\documentclass[journal]{IEEEtran}
\usepackage{amsmath,amssymb,amsfonts}
\usepackage{algorithmic}
\usepackage{graphicx}
\usepackage{textcomp}
\usepackage[table,xcdraw]{xcolor}
\usepackage[tight,footnotesize]{subfigure} 
\usepackage{stfloats}
\usepackage{hyperref}
\usepackage{booktabs}
\usepackage[numbers,sort]{natbib}
\usepackage{caption}
\usepackage[nolist]{acronym} 
\acrodef{DNS}{Domain Name System}
\acrodef{PII}{Personal Identifiable Information}
\acrodef{PIS}{Personal Information System}
\acrodef{PMF}{Probability Mass Function}
\acrodef{SDC}{Statistical Disclosure Control} 
\acrodef{LBS}{Location-Based Systems}
\acrodef{RS}{Recommender Systems}
\acrodef{PET}{Privacy-Enhancing Technologies}
\acrodef{DoT}{DNS over TLS}
\acrodef{DoH}{DNS over HTTPS}
\acrodef{DoQ}{DNS over QUIC}
\acrodef{DPT}{Data Perturbation Techniques}
\acrodef{TTPs}{Trusted Third Parties}
\acrodef{PIR}{Private Information Retrieval}
\acrodef{ODNS}{Oblivious DNS}
\acrodef{ODoH}{Oblivious DNS over HTTPS}
\acrodef{KL}{Kullback-Leibler}
\acrodef{IT}{Information Theory}
\acrodef{NQA}{Never Query Alone}
\usepackage{mathrsfs}
\newcommand{\sX}{\mathscr{X}}  
      
\DeclareMathOperator{\oD}{D}
\DeclareMathOperator{\oH}{H}
\newcommand{\cR}{\mathcal{R}}
\DeclareMathOperator*{\argmin}{arg\,min}

\usepackage{bm}

\usepackage[T1]{fontenc}
\makeatletter
\DeclareSymbolFont{NewLetters}{T1}{times}{m}{it}
\AtBeginDocument{
  \DeclareMathVersion{bold}
  \SetSymbolFont{operators}{bold}{T1}{times}{b}{n}
  \SetSymbolFont{NewLetters}{bold}{T1}{times}{b}{it}
  \SetMathAlphabet{\mathrm}{bold}{T1}{times}{b}{n}
  \SetMathAlphabet{\mathit}{bold}{T1}{times}{b}{it}
  \SetMathAlphabet{\mathbf}{bold}{T1}{times}{b}{n}
  \SetMathAlphabet{\mathtt}{bold}{OT1}{pcr}{b}{n}
  \SetSymbolFont{symbols}{bold}{OMS}{cmsy}{b}{n}
  \renewcommand\boldmath{\@nomath\boldmath\mathversion{bold}}
}
\makeatother

\def\BibTeX{{\rm B\kern-.05em{\sc i\kern-.025em b}\kern-.08em
    T\kern-.1667em\lower.7ex\hbox{E}\kern-.125emX}}

\begin{document}
\title{DNS Query Forgery: A Client-Side Defense Against Mobile App Traffic Profiling}

\author{
\IEEEauthorblockN{Andrea Jimenez-Berenguel\textsuperscript{1}, César Gil\textsuperscript{2}, Carlos Garcia-Rubio\textsuperscript{1}, Jordi Forné\textsuperscript{2}, Celeste Campo\textsuperscript{1}}

\IEEEauthorblockA{\textsuperscript{1}Dept. of Telematic Engineering, Universidad Carlos III de Madrid, Leganés (Madrid), Spain}

\IEEEauthorblockA{\textsuperscript{2}Dept. of Telematic Engineering, Universitat Politècnica de Catalunya, Barcelona, Spain}
\IEEEcompsocitemizethanks{\IEEEcompsocthanksitem Corresponding author: Andrea Jimenez-Berenguel (andrejim@pa.uc3m.es).}
}

\maketitle


\begin{abstract}
Mobile applications continuously generate DNS queries that can reveal sensitive user behavioral patterns even when communications are encrypted. This paper presents a privacy enhancement framework based on query forgery to protect users against profiling attempts that leverage these background communications. We first mathematically model user profiles as probability distributions over interest categories derived from mobile application traffic. We then evaluate three query forgery strategies—uniform sampling, TrackMeNot-based generation, and an optimized approach that minimizes Kullback-Leibler divergence—to quantify their effectiveness in obfuscating user profiles. Then we create a synthetic dataset comprising 1,000 user traces constructed from real mobile application traffic and we extract the user profiles based on DNS traffic. Our evaluation reveals that a 50\% privacy improvement is achievable with less than 20\% traffic overhead when using our approach, while achieving 100\% privacy protection requires approximately 40-60\% additional traffic. We further propose a modular system architecture for practical implementation of our protection mechanisms on mobile devices. This work offers a client-side privacy solution that operates without third-party trust requirements, empowering individual users to defend against traffic analysis without compromising application functionality.
\end{abstract}




\section{Introduction}

Over the past decade, mobile application usage has grown substantially due to improved device capabilities, increased high-data content, and enhanced network performance~\cite{ericsson2024mobilityreport}. As users interact with these applications (hereafter abbreviated as apps), they explicitly and implicitly disclose personal information. This information is typically utilized by~\ac{PIS}, such as \ac{LBS} and \ac{RS}, which tailor services based on user data. However, the very process of personalization introduces inherent privacy risks. As personal data flows into countless online services, unintended parties—including cybercriminals and network observers—gain more ways to target users and their assets. In response to these privacy concerns, \ac{PET} have been in development since the late 20th century, aiming to protect user information in data-driven environments.

The vulnerabilities of user profiling, traditionally associated with foreground interactions, is not limited to direct app-server communications. Background interactions as \ac{DNS} traffic also enable user profiling. Every time a user loads a webpage, or sends a message, \ac{DNS} is the first checkpoint on the path as it is responsible for translating human-readable domain names into machine-readable IP addresses. The \ac{DNS} queries expose sensitive data that may be exploited for profiling purposes.

Network observers such as eavesdroppers or DNS resolvers can infer user behavior from the domain patterns. Despite of the development of encrypted DNS protocols such as \ac{DoT}, \ac{DoH}, and \ac{DoQ} their overall adoption remains relatively low as stated in the study~\cite{Lyu2022dns}. DNS queries continue to be transmitted in clear text over port 53 (Do53) as demonstrated in the analysis conducted by~\cite{Campo2024dns}.

Although encrypted DNS communications between clients and resolvers prevents eavesdropping, a malicious or compromised resolver can still violate user privacy. Moreover, most encrypted‐DNS traffic is handled by a few major providers such as Google and Cloudflare~\footnote{https://dnscrypt.info/public-servers/}. Such centralization could lead to data monopolization that ultimately harms clients’ privacy. There are emerging DNS Privacy‐preserving solutions such as~\cite{Bhat2019pir,Arana2021nqa,Schmitt2019odns,Singanamalla2020odoh} to address this vulnerability.

As shown in our analysis, domain names from DNS queries within a fixed time window readily expose user activity patterns, even with encrypted communications. This fundamental vulnerability exists wherever DNS observers can see queried domains. Our main contribution addresses this privacy risk by applying \ac{DPT} to DNS traffic through query forgery. While previous work like~\cite{gil2025privacy} has applied these techniques to foreground interactions in location-based systems, we extend them to background DNS communications. Our approach mixes genuine DNS queries with strategically generated false ones, creating an obfuscated view of user behavior. Unlike traditional DPT applications that balance personalization utility against privacy, our DNS implementation only incurs network overhead costs. This overhead represents a reasonable trade-off for the significant privacy protection gained against profiling attempts based on DNS traffic analysis.

In this context, this paper makes the following main contributions. First, we present a mathematical model of user profiles derived from DNS queries and we apply \ac{DPT} via query forgery to enhance DNS privacy. Second, we evaluate three false-query strategies—uniform sampling, a TrackMeNot-inspired generator~\cite{howe2009lessons}, and a \ac{KL}-divergence–minimizing optimizer~\cite{rebollo2010optimized}. Third, we build a synthetic dataset by mapping a real dataset traffic onto 1,000 users and we profile these users with our proposed method based on DNS traffic. Fourth, through theoretical analysis and experimental evaluation, we show that query forgery substantially reduces profiling accuracy with minimal performance degradation, and we quantify the trade-off between added network overhead and achieved privacy gains. Finally, we propose a modular system architecture for a potential implementation of our DNS privacy model. 

The remainder of this paper is organized as follows. Section~\ref{sec:State of the Art} presents a comprehensive review of the state of the art. In Section~\ref{sec:Formal Problem Statement}, we formally define the problem of DNS-based user profiling. Section~\ref{sec:DNS Query Forgery Against Profiling} introduces our proposed DNS query forgery mechanism to enhance user privacy against profiling based on DNS traffic. Section~\ref{sec:Users_Dataset} describes the creation of our synthetic dataset and outlines the methodology for generating user traces and deriving DNS-based user profiles. In Section~\ref{sec:Evaluation}, we present our experimental evaluation, analyzing the trade-off between privacy enhancement and traffic overhead, and discuss the results obtained from applying our method to 1,000 synthetic users. We propose a practical adaptation of query forgery for mobile apps in Section~\ref{sec:Practical Adaptation of Query Forgery for Mobile Apps}. Finally, Section~\ref{sec:Conclusion} concludes the paper and outlines directions for future research.

\section{State of the Art}
\label{sec:State of the Art}
\noindent


In this section, we review three key areas of literature. First, we examine studies on user profiling through network traffic analysis. Next, we explore \ac{PET}s and their real-world applications. Finally, we review existing works that apply PETs to mitigate privacy vulnerabilities in DNS communications. This review frames our contribution within the broader privacy protection landscape.



\subsection{User profiling}
\label{sec:State of the Art:User profiling}
\noindent

User profiling through network traffic analysis involves constructing behavioral patterns of users based on their digital activities. This subsection reviews various methodologies used to extract user profiles from network traffic parameters. Each approach leverages distinct aspects of network communications to reveal behavioral patterns and user interests. We categorize the literature into three main groups: profiling based on HTTP/HTTPS parameters, DNS parameters, and broader network metadata, as summarized in Table~\ref{tab:user_profiling_summary}.


\begin{table}[!t]
\centering
\resizebox{\columnwidth}{!}{%
\begin{tabular}{lll}
\toprule
\textbf{Profiling Parameter Type} & \textbf{References} & \textbf{Applications} \\
\midrule
HTTP/HTTPS parameters & \cite{Gonzalez2016userprofiling,Gonzalez2021userprofiling,Park2018userprofiling,Gao2019userprofiling} & Personality inference, academic classification \\
\midrule
DNS parameters & \cite{Shaman2019userprofiling,Lyu2023profiling} & Behavior profiling \\
\midrule
Network metadata parameters & \cite{Alotibi2016userprofiling,Li2022userprofiling} & Activity pattern recognition \\
\bottomrule
\end{tabular}%
}
\caption{User profiling approaches by parameter type, references, and applications.}
\label{tab:user_profiling_summary}
\end{table}

Several studies have exploited the characteristics of HTTP/HTTPS traffic to build user profiles. Although HTTPS makes profiling more difficult, it does not eradicate it. Gonzalez et al.~\cite{Gonzalez2016userprofiling,Gonzalez2021userprofiling} demonstrated that user profiling is possible despite encryption. From an eavesdropper's perspective, in~\cite{Gonzalez2016userprofiling}, they utilized the URL a user visits, which can be obtained from the Server Name Indication (SNI) in the TLS handshake's client\_hello message, and web fingerprinting techniques to construct user profiles. From a network observer's perspective, in~\cite{Gonzalez2021userprofiling}, they analyzed sequences of hostnames visited by users within a predefined period of time. Using a custom Chrome extension to collect browsing data, they applied natural language processing algorithms to infer relevant topics from accessed websites, effectively creating user profiles despite encryption protections.

Building on HTTP/HTTPS-based profiling approaches, Park et al.~\cite{Park2018userprofiling} conducted a comprehensive investigation into four distinct profiling scenarios: (i) profiling based on timestamps; (ii) profiling based on HTTP headers; (iii) profiling based on domain names, assuming interpretable topical categories of URLs; and (iv) profiling based on page content, noting its inapplicability to HTTPS traffic where URLs are encrypted and only domain names remain visible. Their research used a proprietary dataset of mobile network traffic from 61 Spanish participants. By analyzing HTTP(S) traffic patterns of the users, they modeled personality traits, shopping interests, and demographic characteristics.

Gao et al.~\cite{Gao2019userprofiling} integrated HTTP(S) parameters with network access records in a campus environment to enhance user profiling capabilities. Their methodology first extracted fundamental identifiers (MAC address, login/logout times, device location) from network access records, then enriched this data by analyzing HTTP(S) packets for additional identifiers like destination IP addresses. Using these parameters, they built a classifier based on Back Propagation Neural Networks (BPNN) to distinguish between different user types and predict academic disciplines.

Several researchers have explored DNS traffic as a rich source of information for user profiling. Shaman et al.~\cite{Shaman2019userprofiling} proposed a method for user identification and behavior profiling using DNS information. They collected a dataset from 23 users on the Plymouth University network, filtered users based on MAC/IP address mappings, and identified apps using reverse DNS queries. The authors employed a gradient boosting machine learning algorithm to create user profiles based on features such as date and time, destination IP address, and DNS queries. 

Lyu et al.~\cite{Lyu2023profiling} conducted an analysis of unencrypted DNS traffic collected over one month from both a university campus and a government research institute. Their research identified distinctive behavioral patterns among various DNS asset types (recursive resolvers, authoritative name servers, and mixed DNS servers). By capturing normal DNS activity patterns and detecting anomalies indicative of security issues, they demonstrated DNS traffic's dual utility for both profiling and security monitoring. Their implementation of an unsupervised machine learning algorithm to classify over 100 DNS assets based on network, functional, and service characteristics further established DNS behavior profiling as a valuable tool for automated security management.

Beyond HTTP/HTTPS and DNS parameters, broader network metadata provides valuable insights for user profiling without requiring access to communication content. Alotibi et al.~\cite{Alotibi2016userprofiling} developed user behavioral profiles based on network metadata parameters such as connection type, duration, number of packets, and packet size. Their research utilized custom-collected network metadata from 27 participants with static IP addresses, providing ground truth for their analysis. After filtering the traffic to focus on user interactions with popular apps (Google, YouTube, Skype, Facebook, etc.), their methodology achieved remarkable identification accuracies. Their results demonstrate network metadata's effectiveness for forensic investigations targeting insider threats.

Li et al.~\cite{Li2022userprofiling} analyzed user activity sequences using real-world data from a Shanghai ISP. Their methodology constructed user profiles from network access records containing user IDs, timestamps, and connection metadata. Their approach used the ISP's systematic classification of apps into categories (games, shopping, education) and observed that each user's records followed a power-law distribution. By segmenting app usage traces into time windows and applying a probabilistic topic model, they successfully inferred users' cyber activities. Their research conclusively demonstrated that digital activity patterns could characterize users' daily life both individually and collectively.

In conclusion, previous studies has shown that a variety of traffic network parameters—including HTTP/HTTPS attributes (timestamps, headers, hostnames), DNS characteristics (query patterns, response types), and generic flow metadata (durations, packet sizes)—can be used to reconstruct user behavior even when traffic is encrypted. 



\subsection{Privacy-Enhancing Technology Applications}
\label{sec:State of the Art:PET applications}
\noindent

Numerous \ac{PET}s have been proposed in the literature and applied in real-world scenarios. According to~\cite{parra2015privacy}, a possible classification of \ac{PET}s can be summarized into five groups: (a) basic anti-tracking technologies, (b) approaches based on \ac{TTPs}, (c) collaborative mechanisms, (d) methods based on \ac{PIR} cryptography, and (e) \ac{DPT}s. We concentrate on \ac{DPT}s which aim to obfuscate the data users share with \ac{PIS}. In practice, these techniques are specifically designed to hinder the precise profiling of users by third-party privacy attackers. The paradigmatic example of \ac{DPT} is the transmission of real user data mixed with false data.

Among the five broad PET categories, \ac{DPT}s operate under a zero-trust model regarding third-party entities. Unlike other approaches that rely on trusted intermediaries, \ac{DPT}s treat any third party as a potential privacy threat. This approach implements local privacy (user-side privacy), although it can still be combined with collaborative profiling mechanisms when appropriate.

Another important property of \ac{DPT}s in the context of \ac{PIS}s is the trade-off between cost and benefit. The primary goal of these techniques is to find a balance between system functionality cost (personalization), which depends on data utility, against user privacy protection, which mitigates profiling risk. In DNS traffic between mobile apps and DNS servers—where no personalization occurs—this functionality cost manifests as network overhead rather than reduced service quality.

While DPTs are typically applied to PISs, their principles can be extended to the scenario of DNS traffic generated by mobile apps. This subsection examines the different approaches to data perturbation of the literature, which we categorize based on their mechanism: forgery, suppression, hybrid approaches combining both, and generalization, as summarized in Table~\ref{tab:dataperturbation} with their respective references and applications. 

It is important to clarify that, in all cases, we refer to deterministic perturbations, as opposed to techniques that rely on random perturbations. Notably, all these techniques have analogous counterparts in the field of \ac{SDC}, although the object of protection differs. While the studies analyzed in this section focus on protecting a user profile—typically modeled as a \ac{PMF}—\ac{SDC} techniques are designed to safeguard an entire database of records. 

\begin{table}[!t]
\centering
\resizebox{\columnwidth}{!}{%
\begin{tabular}{lll}
\hline
Data-perturbation techniques & References & Applications \\
\hline
$\bullet$ Forgery & \cite{elovici2002newa,elovici2002newb,elovici2005enhancing,elovici2006cluster,ye2009noise,domingo2009h,howe2009lessons,rebollo2010optimized,puglisi2015content} & PWS, PIR, RSs \\
$\bullet$ Supression & \cite{parra2010privacy,parra2012optimal,parra2012privacy} & PWS, RSs \\
$\bullet$ Both & \cite{parra2011privacy,parra2014optimal} & RSs \\
$\bullet$ Generalization & \cite{gil2025privacy} & LBS \\
\hline
\end{tabular}%
}
\caption{\ac{DPT}s, references, and applications.}
\label{tab:dataperturbation}
\end{table}

Query forgery techniques involve adding false queries to genuine ones. This approach allows users to protect themselves from precise profiling by privacy attackers while avoiding the need to rely on third parties.

Several significant query forgery–based proposals had emerged in the literature. The private web browsing system known as PRAW~\cite{elovici2002newa,elovici2002newb,elovici2005enhancing,elovici2006cluster} complicated user profiling by generating false browsing traces when users accessed the web through a shared login session. Similarly, in~\cite{ye2009noise} the authors presented a query injector that generated false queries with probabilities complementary to real ones. That approach assumed that the proportions of real and false queries remained inaccessible to an adversary and were only available on the user side.

A software implementation of query forgery was GooPir~\cite{domingo2009h}. GooPir operated by sending batches of both genuine and false keywords to a web search engine. The selection of false keywords was based on usage frequency similar to that of genuine ones, making profiling attacks more challenging. However, in~\cite{balsa2012ob} the authors highlighted that this strategy could be vulnerable to correlation attacks between keywords across different batches.

Similarly, TrackMeNot~\cite{howe2009lessons} was a web browsing plugin that implemented query forgery using various strategies. False queries were generated through a continuously updated keyword dictionary sourced from diverse information channels. The transmission strategy of false queries to the server could either mimic human behavior through bursts or be set at predefined time intervals. As with the previous proposal,~\cite{chow2009faking} argued that certain semantic or timing‐based attacks on false queries could lead to potential inference of real queries, thus exposing TrackMeNot users.

A fundamental drawback of adding false queries was the implicit traffic overhead it generated. Addressing this challenge required balancing privacy and overhead, a scenario studied in~\cite{rebollo2010optimized} within the field of \ac{PIR}. In that work, the authors presented a mathematical model to achieve an optimal trade‐off between the rate of falsified queries and user privacy.  Later, in~\cite{puglisi2015content}, researchers investigated and validated tag forgery in real‐world scenarios within content‐based \ac{RS}s.

Data suppression was a fully viable and conceptually straightforward DPT, representing the opposite approach to adding activity to a user profile.  Suppression had been validated in various scenarios. In~\cite{parra2010privacy}, applied to the Semantic Web, a limited privacy improvement was achieved through a tag removal process, incurring resource costs that traded off with semantic degradation. The same objective was analytically explored via convex optimization in~\cite{parra2012optimal}. Shannon entropy of the perturbed profile and the proportion of tags the user was willing to remove served as the respective privacy and utility metrics for studying the optimal balance. Finally, in~\cite{parra2012privacy}, parental control and resource recommendation were presented as application scenarios. The evaluation of suppression‐based perturbation considered costs resulting from data degradation and the accuracy of predefined parental control policies, offering an insightful perspective.

A combined approach involving both forgery and suppression was also applied to personalized RSs (e.g., Amazon, Spotify, Netflix). Essentially, this strategy, investigated in~\cite{parra2011privacy}, enabled users to submit false ratings and/or withhold ratings for items of interest. A closed‐form solution to the problem of optimal and simultaneous forgery and suppression of real‐world ratings was presented in~\cite{parra2014optimal}.

More recently, in~\cite{gil2025privacy} the authors proposed and evaluated a real‐world data perturbation strategy based on the generalization of interest categories arranged in a hierarchical taxonomy with varying depth levels. That study proved effective in systems utilizing hierarchical semantic taxonomies or in LBSs, where POI coordinates could be recursively categorized within a properly partitioned area of interest. As in previous studies, and among other properties of the convex optimization problem modeling the privacy–utility trade‐off, the authors introduced a critical ratio as a measure of the maximum generalization rate beyond which privacy could not be further improved.  

In conclusion, \ac{DPT}s present various methodologies for enhancing user privacy protection. Query forgery mechanisms, as implemented in systems like PRAW, GooPir, and TrackMeNot, function by strategically mixing false queries with genuine ones to obfuscate user profiles. Data suppression techniques operate on the contrary principle, selectively removing sensitive elements from user profiles to limit information disclosure. Hybrid approaches combine both falsification and suppression strategies to optimize privacy protection, while generalization techniques organize sensitive data into hierarchical taxonomies that balance utility with privacy. These different approaches demonstrate the versatility of \ac{DPT}s in addressing privacy challenges across various app domains.



\subsection{Applications of PETs to DNS}
\label{sec:State of the Art:Applications of PETs to DNS}
\noindent


Beyond \ac{DPT}, researchers have also explored \ac{PIR} for DNS privacy. Bhat et al.~\cite{Bhat2019pir} propose an information‐theoretically perfect, single‐database PIR scheme for private DNS resolution, and more recently Zhou et al.~\cite{Zhou2024pir} introduce PIANO, a highly practical, single‐server PIR with sublinear server work that scales to 100 GB DNS‐sized datasets. However, DNS records are frequently updated, which contradicts a fundamental assumption of PIANO that requires a static database. Moreover, PIR privacy requires also server support so they cannot be deployed purely on the client side. This makes private DNS query particularly challenging to implement with PIR-based approaches.


In the same spirit of DNS privacy, Arana et al.~\cite{Arana2021nqa} propose \ac{NQA}, a cooperative routing strategy in which users forward their DNS queries through their neighbors, thereby diluting an attacker’s ability to link a query to its source. \ac{NQA} requires cooperation among multiple clients. 

Researchers have also explored privacy-enhancing solutions specifically for DNS. Schmitt et al.~\cite{Schmitt2019odns} propose \ac{ODNS}, which decouples client identity from queries by encrypting requests that recursive resolvers forward to \ac{ODNS} resolvers. Building on this concept, Singanamalla et al.~\cite{Singanamalla2020odoh} developed \ac{ODoH}, adding HTTPS transport to \ac{ODNS}. Cloudflare has implemented \ac{ODoH} in their public DNS resolver~\footnote{https://developers.cloudflare.com/1.1.1.1/encryption/oblivious-dns-over-https/}, and the protocol is being standardized through an IETF draft co-authored by Cloudflare and Apple~\cite{rfc9230}.





In conclusion, building upon the aforementioned works, our paper aims to demonstrate that the domain names queried over a fixed time window leaks rich user‐level behavior patterns. As long as eavesdroppers or network observers sees those domain names, user activity remains fundamentally exposed. To counter this we apply DPT via query forgery not only because it empowers users to shield themselves without relying on systems that require server-side deployments, but also because it operates under a zero‐trust model for any external entity and strikes a practical balance between system‐functionality cost and user‐privacy protection. 



\section{Formal Problem Statement} 
\label{sec:Formal Problem Statement}
\noindent

Generally, individual private data generated and transacted between users and PISs over communication networks can be represented as sequences of random variables. In this work, we adopt this representation for DNS traffic, specifically DNS queries or parts thereof (e.g., tuples consisting of the app issuing a query, the queried domain, and the timestamp) generated by user devices through various installed mobile apps. Ultimately, these sequences can assume values in a finite, common, and reduced alphabet of categories, which we define as the set $\sX = \{1,\ldots,n\}$ for some integer $n \ge 2$.

Assuming that these random variables are independent and identically distributed, we mathematically model a user's profile using a \ac{PMF} over the distribution of these variables. We profile each user based on the percentage of DNS queries generated by each app in a predefined interval. The primary advantage of \ac{PMF}, widely accepted in the privacy literature~\cite{xu2007privacy,toubiana2010adnostic,rebollo2010optimized,fredrikson2010repriv,parra2017myadchoices}, is its ability to efficiently aggregate large amounts of individual user data and present it as a histogram of relative frequencies across predefined interest categories. Consequently, user profiles based on discrete probability distributions are well-suited for numerical computation and widely applicable in privacy metrics.

Our goal in this formulation is not to identify specific individuals or extract personal details, but rather to formalize how DNS query patterns inherently reveal user behavior signatures. We aim to establish the mathematical foundation for understanding the fundamental privacy vulnerability that exists whenever DNS traffic can be monitored by third parties, regardless of how that information might be exploited.


Considering these aspects, we define $q$ as the distribution of genuine DNS queries from a user, reflecting their interests (e.g., entertainment, culture, etc.) based on the DNS traffic generated by mobile apps installed on their device. Similarly, we define $r$ as the distribution of the user's false queries and $p$ as a reference distribution, which may correspond to the population distribution or the average distribution of a user group. In profiling terms, $q$, $r$, and $p$ represent the user's real profile, false profile, and reference profile, respectively.

Finally, as a result of profile mixing, we define $t$ as the user's apparent profile, derived from the combination of the real and false profiles. In this work, $t = (1-\rho)p + \rho r$, a simple deterministic perturbation strategy based on the convex combination of genuine and false queries, where $\rho$ is the false query rate, also known as the perturbation ratio, with values between $0$ and $1$.


With the formal problem statement established, the following section presents the key components of our proposal to enhance privacy protection against user profiling based on DNS traffic by employing DNS query forgery. Ultimately, we aim to determine a distribution of false queries $r$ that optimally obfuscates the real user profile $q$ in terms of privacy and utility, making it indistinguishable to a privacy attacker observing the apparent profile $t$ and leveraging DNS traffic for profiling activities.

\section{DNS Query Forgery Against User Profiling}
 \label{sec:DNS Query Forgery Against Profiling}
 \noindent

\begin{figure*}[!ht]
	\centering
	\includegraphics[width=\textwidth]{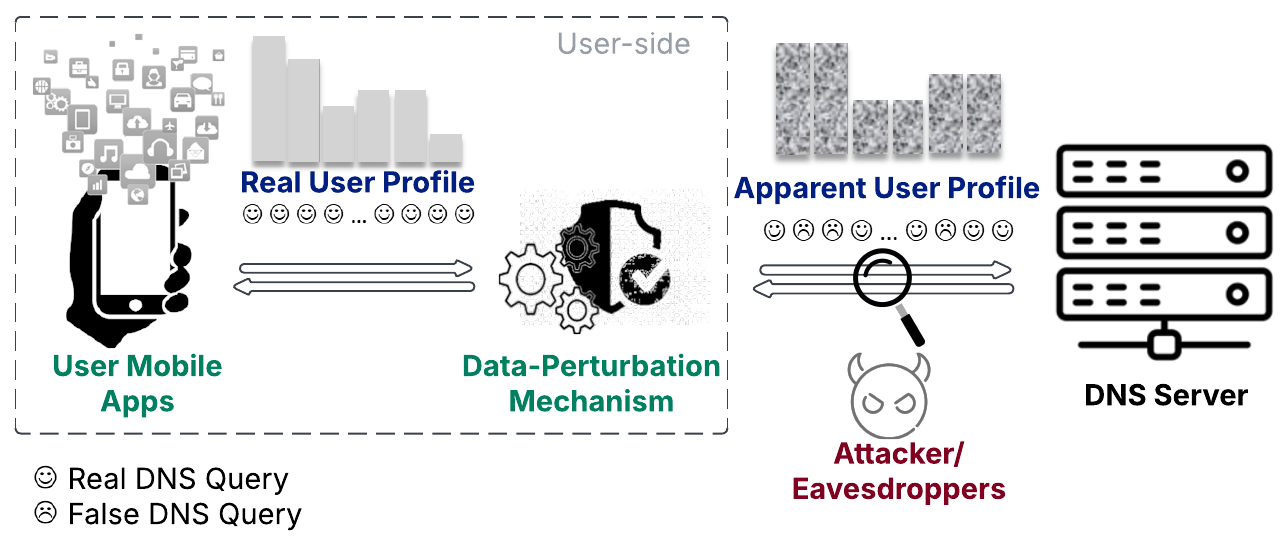}
	\caption{Representation of the secured scenario with the user model, the data-perturbation mechanism and attacker model.}
	\label{fig:diagram_scenario}
\end{figure*}

In this section, we present our proposal for enhancing user privacy against profiling based on DNS traffic generated by mobile apps installed on their devices. Fig.~\ref{fig:diagram_scenario} illustrates the diagram of the proposed scenario. First, we define the user and adversary models that we assume. Next, we introduce the metrics that will be used to evaluate the obfuscation strategy forming the basis of our privacy-utility trade-off model. Finally, we provide a numerical example to illustrate the proposed privacy model. 


\subsection{User Model}
\label{sec:DNS Query Forgery Against Profiling:User Profile Model}
\noindent

The first component to consider in a security analysis is the entity that a privacy attacker will observe, taking into account the parameters that may be compromised.
As introduced in Section~\ref{sec:Formal Problem Statement}, the scenario considered in this work is based on a stream of DNS queries. At a given instant or over a specific time period, processing and aggregating part of the information contained in these queries allows the construction of a user profile in the form of a discrete histogram of relative frequencies, summarizing the user's preferences across a finite set of predefined interest categories. A straightforward definition of these categories can be derived from mapping the mobile apps used by the user on their device, an association inferred from the traces of DNS queries.

This user model, widely studied and employed in the field of PISs, enables us to define and delineate the profiling attack performed by an adversary. It is important to reiterate that user activity profiling is a primary concern in such systems.

In general, profiling is a method used to identify and characterize individuals by generating and applying profiles. However, as discussed in the literature~\cite{hildebrandt2005descriptive,hildebrandt2008profiling}, user identification can be understood from two distinct perspectives: as \textit{individuation}, referring to the revelation of an individual's unique attributes, and as \textit{classification}, which involves categorizing an individual as a member of a group.

This duality in profiling usage implies the creation of both individual and group profiles. For instance, PISs are commonly characterized by individual profiling, which personalizes services based on each user's specific interests. Conversely, other systems seek to adapt a group profile to users who may not have directly contributed to that profile.

In this work, we focus on user individuation, adapting the adversary model and privacy metrics to this specific profiling activity. Ultimately, the real and apparent user profiles, denoted as $q$ and $t$, respectively, will be the targets of the adversary's profiling, whose model we define in the following section.
 
\subsection{Attacker Model}
\label{sec:DNS Query Forgery Against Profiling:Adversary Model}
\noindent

The level of privacy provided by a PET directly depends on the assumptions we make about the adversary. For this reason, evaluating the effectiveness of a PET requires a proper characterization of the privacy attacker. Clearly, depending on the adversary's properties, a user may implement different techniques, including those reviewed in Section~\ref{sec:State of the Art:PET applications}.

Throughout this work, we consider an adversary capable of accessing the DNS traffic generated by the exchange of information between the various mobile apps installed on a user's device and the servers that resolve DNS queries—an essential requirement for the apps to function properly.

Given this access, we assume an adversary who can filter DNS queries and profile users by inferring their interests through the analysis of the information contained in the traces recorded on DNS servers, following approaches such as those described in~\cite{Campo2024dns}. The technique we propose assumes that a user aims to conceal their bias toward specific categories of interest by perturbing their traffic with false DNS queries. This ensures that the apparent profile $t$, as observed by any attacker, approximates either a uniform profile $u$ or the average user profile $\bar{q}$, while deviating as much as possible from the real profile $q$.

A final but crucial assumption regarding the privacy attacker is their inability to estimate a user's rate of false queries $\rho$. This is based on the premise that the adversary lacks knowledge of whether the user employs the proposed privacy strategy.

\subsection{Metrics}
\label{sec:DNS Query Forgery Against Profiling:Metrics}
\noindent

We dedicate this section to justifying and describing the privacy and utility metrics selected for our privacy protection proposal concerning user DNS traffic in mobile apps. For a detailed analysis, these metrics have been extensively studied in~\cite{rebollo2010optimized,parra2014measuring}.

First, we chose to use privacy measures derived from \ac{IT}. Specifically, we handle two key concepts: Shannon entropy and \ac{KL} divergence. For readers unfamiliar with this field, we briefly review both measures below.

The Shannon entropy of a discrete random variable with PMF $q$ taking values in the set $\sX = {1,\ldots,n}$ is a measure of the uncertainty of this random variable and is defined as

	\begin{equation}\label{eqn:entropy}
		\oH(q) = \sum_{i}  q_{i} \log_{b}(q_{i}).
	\end{equation}	

where $b$~\footnote{Common values of $b$ are
$2$, $e$ and $10$. In those cases, the units of entropy are bit, nat and dit, respectively} is the base of the logarithm used. However, all bases produce equivalent optimization objectives.

Similarly, the \ac{KL} divergence between two discrete random variables with PMFs $q$ and $p$ is a measure of their divergence, also referred to as relative entropy, as it generalizes the Shannon entropy of one distribution with respect to another. It is defined as
	\begin{equation}\label{eqn:relative_entropy}
		\oD(q||p) = \sum_{i} q_{i} \log_{b}(\frac{q_{i}}{p_{i}}).
	\end{equation}	

It is worth noting that Shannon entropy can be considered a special case of \ac{KL} divergence when the reference distribution $p$ is the uniform distribution $u$, i.e., $p=u$.

With these notions in place, we define the privacy risk function $\cR$ as the divergence between the user's apparent profile $t$ and the reference profile $p$, that is,

    \begin{equation}\label{eqn:risk_function}
	\cR(\rho) = \oD((1-\rho)q + \rho r||p), \rho \in   [0,1].
    \end{equation}	
  
Recall that the apparent profile $t$ results from applying a simple perturbation strategy based on a convex combination equivalent to mixing the user’s real profile $q$ with the false profile $r$ in a proportion $\rho$, which we refer to as the perturbation rate.

At this point, we address the selection of the user’s false profile $r$ as a central element of the false DNS query mechanism investigated in this work. To this end, we consider three variants for shaping this discrete distribution.

First, the simplest option consists of diluting the user’s real profile with the uniform distribution $u$, applying the same number of false DNS queries, $1/n$, to each of its $n$ components, which correspond to categories in our case.

The second option is based on the well-known TrackMeNot mechanism~\cite{howe2009lessons}. For practical and simplification purposes, we assume that the false query distribution proposed by this PET is the average of the distribution of a set of users, which we denote as $\bar{q}$.

Finally, we consider an optimized option based on the proposed metrics. In essence, the false query distribution $r$ results from optimizing the convex problem formulated as minimizing \ac{KL} divergence in the unit simplex $\Delta_{r}$ ~\cite{rebollo2010optimized}, i.e.,
    \begin{equation}\label{eqn:optimal_forgery_distribution}
	r^{*} =  \argmin_{r \in \Delta_{r}} \oD((1-\rho)q + \rho r||p).
    \end{equation}

It is important to emphasize that a theoretical analysis of the properties of the function $\cR$ is initially based on the assumption that the distributions $q$ and $p$, understood as probabilities, are strictly positive:

    \begin{equation}
    \label{eqn:assumption1}
	q_{i}, p_{i} > 0 \quad	\text{for all} \quad i=1,\ldots,n.
    \end{equation}	
    
However, in our work, we may refer to continuity arguments to relax this assumption. Moreover, without loss of generality, we assume that

    \begin{equation}
    \label{eqn:assumption2}	
	\frac{q_{i}}{p_{i}} \leq \cdots \leq \frac{q_{n}}{p_{n}} \quad \text{for all} \quad i=1,\ldots,n.
\end{equation}

We highlight to the reader the initial and final values of the privacy risk function, $\cR(0) = \oD(q||p)$ and, in the case of the optimized option, $\cR(1) = 0$. Analyzing the behavior of $\cR$ for intermediate values of $\rho$ when \ac{KL} divergence is minimized in the unit simplex $\Delta_r$ reveals important properties, such as monotonicity and convexity, the existence of a critical perturbation rate $\rho_{crit}$ beyond which privacy is maximized ($\cR(\rho) = 0$ for $\rho \ge \rho_{crit}$), and the existence of an optimal and closed-form solution. The theoretical value of this critical rate is expressed as

    \begin{equation}
    \label{eqn:critical_rate}	
        \rho_{crit} = 1-\frac{q_n}{p_n}
    \end{equation}

In general, the user's knowledge of the reference distribution $p$ determines whether the appropriate metric is divergence or entropy. Table~\ref{tbl:DNS_query_forgery_summary} summarizes the selected variants for the false DNS query mechanism that forms the core of our research proposal.

\begin{table}
	\centering
	\caption{Summary of the three DNS query forgery strategies we investigated in this paper.}
	\label{tbl:DNS_query_forgery_summary}
	\begin{tabular}{ll}
		\toprule
		Mechanism & Distribution  \\ \midrule 
		Uniform (UNF) & $u$ (uniform distribution) \\
		TMN-based (TMN)~\cite{howe2009lessons} & $\bar{q}$ (TMN distribution) \\
		Optimized (OPT)~\cite{rebollo2010optimized} & $r^{*} = \argmin_{r \in \Delta} \oD(t||p)$ \\ \bottomrule
	\end{tabular}
\end{table}

Finally, as a utility metric for our proposal, we directly consider the perturbation rate $\rho$. Intuitively, a higher false query rate leads to greater traffic overhead and a more significant degradation of the user's original profile, making precise profiling— a privacy threat— more challenging. We understand that an increase in DNS traffic reduces the user's quality of experience with their mobile apps.

\subsection{Numerical Example}
\label{sec:DNS Query Forgery Against Profiling:Numerical Example}
\noindent

We dedicate this final section to illustrating the privacy model proposed in this work. To this end, we present some results based on an example that allows the reader to become familiar with the introduced concepts. A more in-depth evaluation of our mechanism in a real-world scenario is presented in Section~\ref{sec:Evaluation}.

Let us consider a scenario where, over a given period and using a standard device, a user's mobile apps have sent 100 queries to a reference information system. These queries can be grouped into a set of five categories summarizing the user's interests, denoted alphabetically as $\{a, b, c, d, f\}$. Assuming that the query frequencies are $(5, 15, 20, 25, 35)$, the user's actual profile with $n=5$ categories is consequently $q=(0.05, 0.15, 0.20, 0.25, 0.35)$. Additionally, we consider the uniform profile $u=(0.20, 0.20, 0.20, 0.20, 0.20)$ and the population profile $\bar{q}=(0.01, 0.05, 0.15, 0.35, 0.44)$.

Given these parameters, the initial value of the risk function based on \ac{KL} divergence when $p=u$ (equivalent to Shannon entropy) and when $p=\bar{q}$ is the same, i.e., $\cR(0) = 0.1386$. However, the final value varies depending on the perturbation strategy and the metric implemented.

\begin{figure}
	\centering
	\subfigure[Entropy-based risk function]{\includegraphics[width=0.6\linewidth]{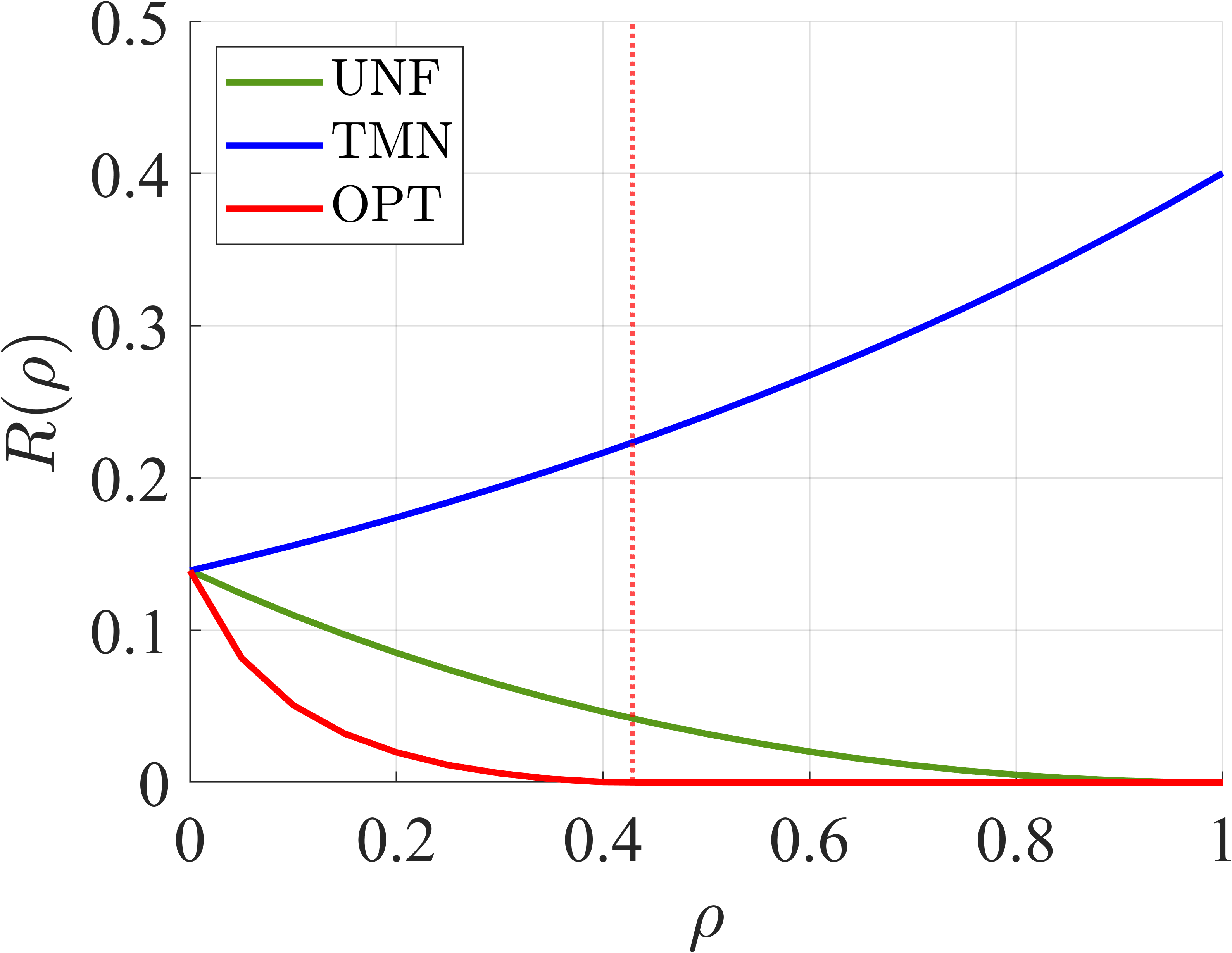}}
	\subfigure[Divergence-based risk function]{\includegraphics[width=0.6\linewidth]{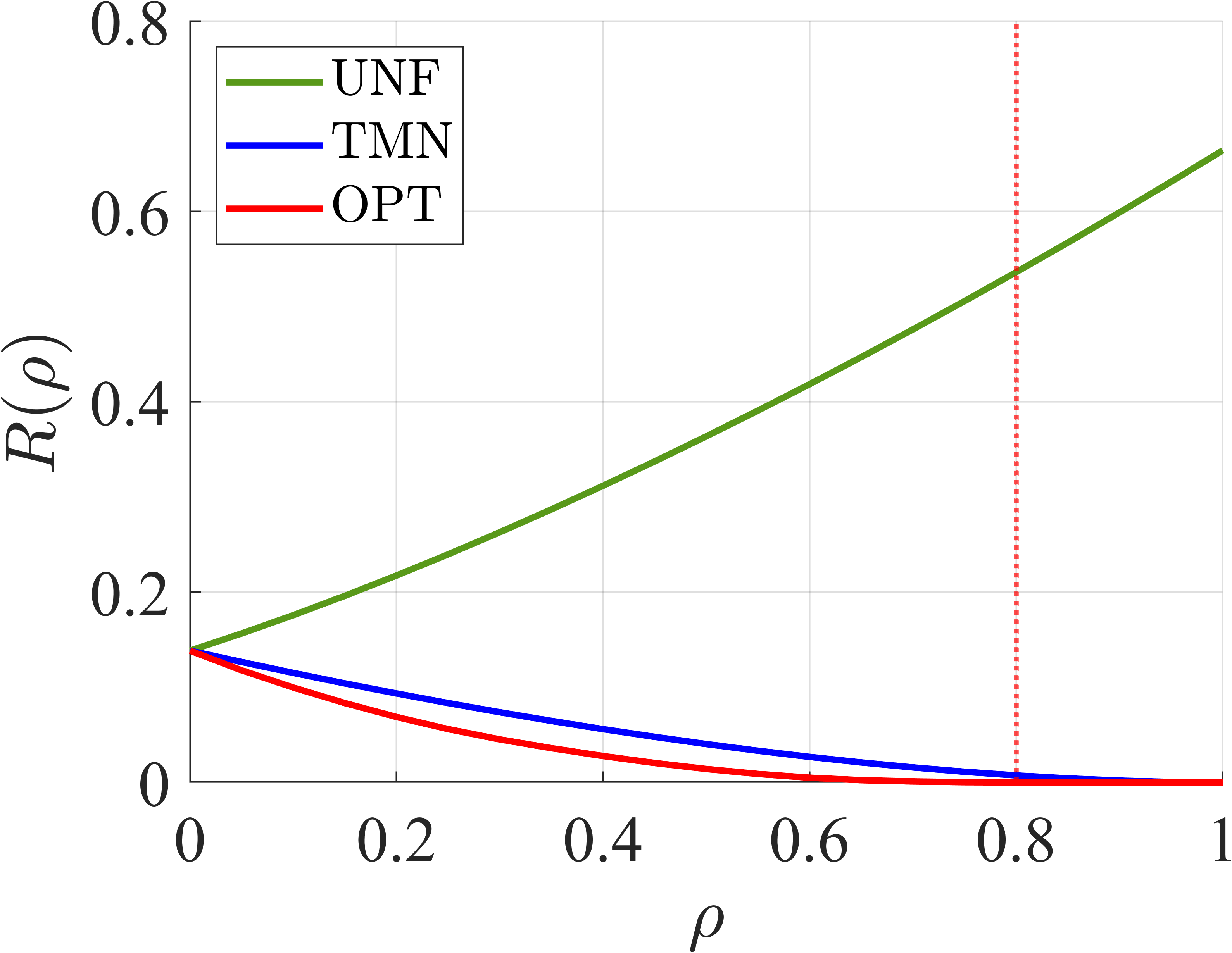}}	
	\caption{Privacy risk $\cR(\rho)$ according to the perturbation ratio $\rho$ for the numerical example for the three perturbation strategies. Denoted by a red dashed vertical line, the $\rho_{crit}$ value.}    
        \label{fig:example_toy_dataset_privacy_risk}
    \end{figure}

In Fig.~\ref{fig:example_toy_dataset_privacy_risk}, we depict the privacy risk function based on entropy and \ac{KL} divergence for this fictitious user as a function of the perturbation rate $\rho$. In both cases, we show the curves corresponding to the three data perturbation mechanisms introduced in Section~\ref{sec:DNS Query Forgery Against Profiling:Metrics}, namely, uniform (UNF), TrackMeNot-based (TMN), and optimized (OPT). The figure also includes the critical perturbation rate $\rho_{crit}$, specific to the optimized mechanism, defined as the minimum rate beyond which the user's privacy risk is null. When using entropy as the metric, $\rho_{crit} = 0.6643$, whereas for \ac{KL} divergence, $\rho_{crit} = 0.8$.

As highlighted in Section~\ref{sec:DNS Query Forgery Against Profiling:Metrics}, the privacy risk function $\cR$ is monotonic and convex in the case of the optimized mechanism. Undoubtedly, this mechanism is superior to suboptimal mechanisms in terms of privacy risk. Intuitively, and as corroborated by the figures, applying a suboptimal mechanism where the reference distribution coincides with the distribution of false queries contradicts the objective of minimizing privacy risk.

\begin{figure}
	\centering
	\subfigure[Entropy-based risk function]{\includegraphics[width=0.85\linewidth]{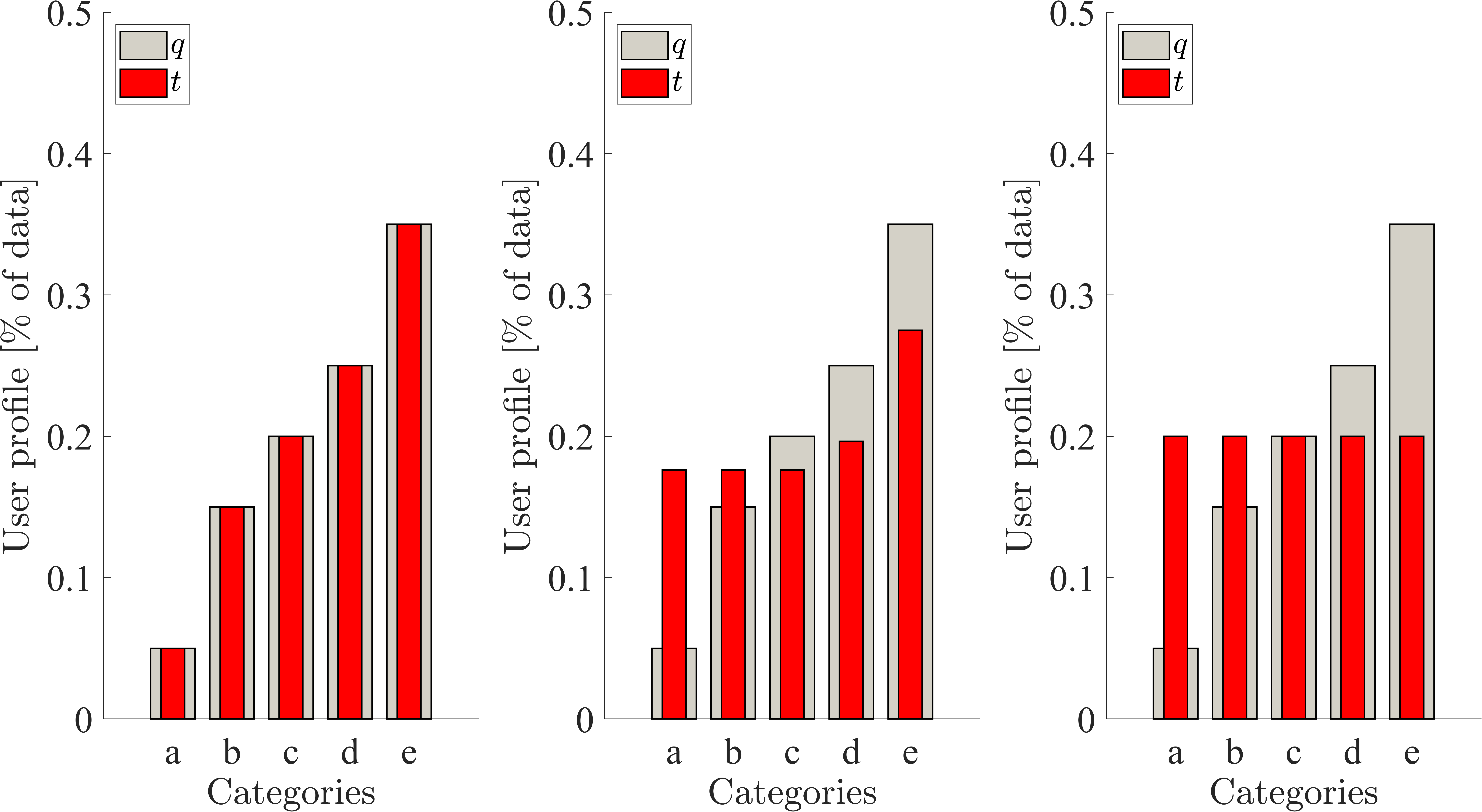}}
	\subfigure[Relative-entropy-based risk function]{\includegraphics[width=0.85\linewidth]{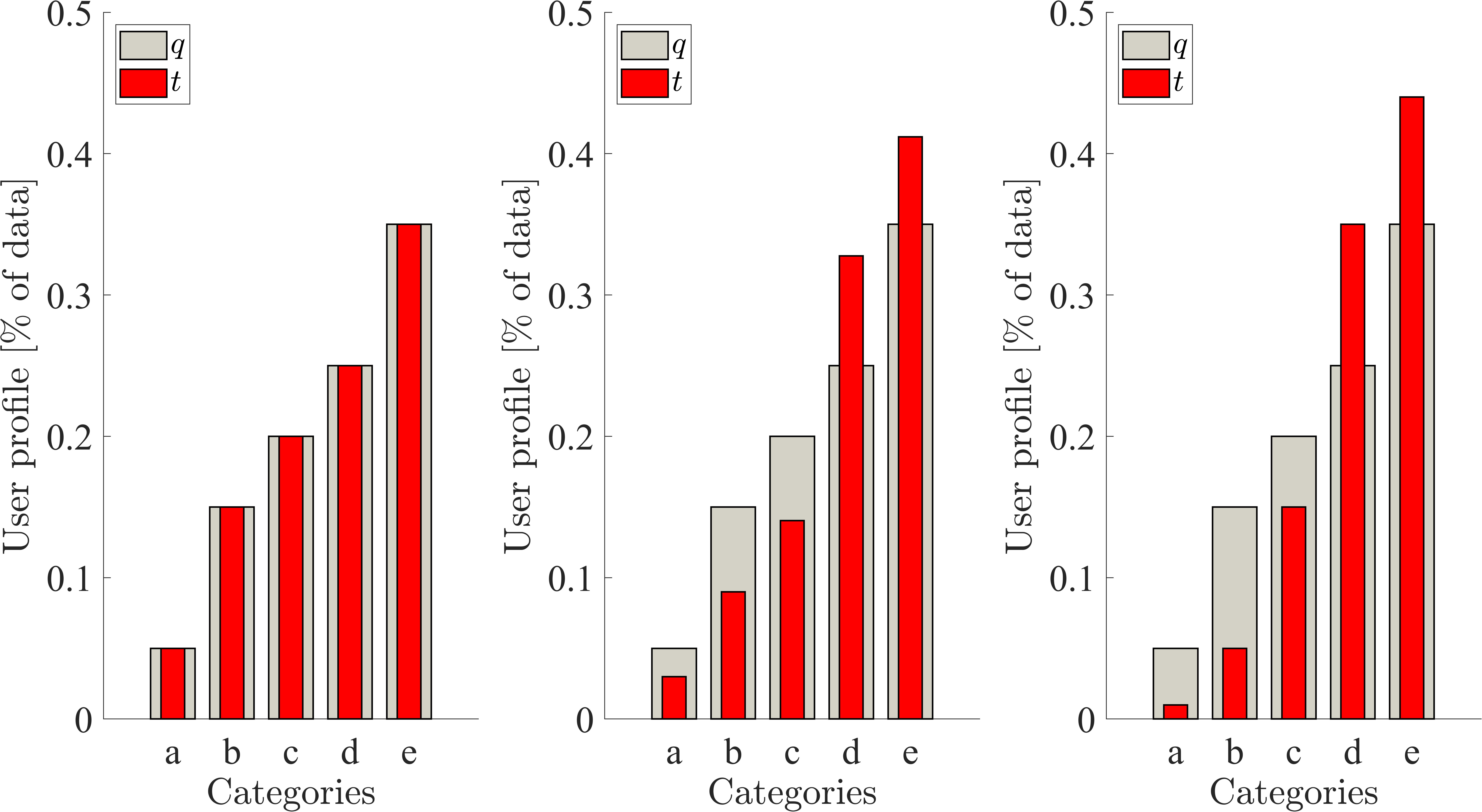}}	
	\caption{Evolution of the real and apparent profile $(q, t)$ for the optimized strategy (oqf) according to different values of the disturbance ratio $\rho = (0, 1/2\rho_{crit}, \rho_{crit})$.}
	\label{fig:example_toy_dataset_histograms}
\end{figure}

Furthermore, Fig.~\ref{fig:example_toy_dataset_histograms} illustrates how the user's profile evolves as the perturbation rate $\rho$ varies between $0$ and $1$ under the optimized mechanism. When $\rho = \rho_{crit}$, the apparent profile $t$ converges to the target or reference profile $p$. This corresponds to the uniform profile when using entropy ($t=u$) and the population profile when using \ac{KL} divergence ($t=\bar{q}$).

We implemented the numerical example and the evaluation presented in Section~\ref{sec:Evaluation} using Matlab (Matlab R2021 9.10.0.1602886 64-bit win64) and executed the computations on an Intel CoreTM i3-2370 CPU at 2.4 GHz, with 4 GB of RAM, running a 64-bit Windows 10 operating system.

To conclude this section, we emphasize that our work presents a comprehensive approach to enhancing user privacy against profiling based on DNS traffic. This is achieved by defining user and adversary models, establishing evaluation metrics to balance privacy and utility, and demonstrating their application through a numerical example. Our model focuses on perturbing user profiles by adding false data to genuine ones, effectively obscuring the real interests of users from potential attackers. To validate the impact of our proposal under real-world conditions, the next section introduces the user dataset and obtains the user profiles employed in the final evaluation. Our dataset consists of 1,000 synthetically generated user traces, a number we consider sufficiently significant to yield robust experimental results. Each trace is composed by mobile app traffic and represents the traffic generated by a user over a time interval.  




\section{Users Dataset}
\label{sec:Users_Dataset}
\noindent








This section describes the selected mobile app traffic dataset used to create our synthetic users, then details the creation of our synthetic dataset composed of mobile app traffic from 1,000 synthetic users. Finally, we explain the process of extracting user profiles based on DNS traffic. These user profiles will subsequently be used to evaluate our proposed privacy model.

We generate a synthetic dataset because there is a notable lack of publicly available datasets that include traffic tied to individual users. Previous studies that analyzed personal traffic~\cite{Alotibi2016userprofiling, Park2018userprofiling, Gao2019userprofiling, Li2022userprofiling, Gonzalez2021userprofiling} do not release their data due to reasons such as ethical concerns regarding user privacy and the risks associated with profiling real users.

The synthetic user generation consist of assigning specific apps and time intervals of traffic of those apps to each user. To create the synthetic user traces we use mobile app traces. Thus, the traffic traces labeled by app provides a controlled environment in which we have a priori knowledge of the active app at any given time, ensuring reliable labeling of traffic per user. 

In the literature there are several datasets of mobile apps in Packet Captured (PCAP) format files. The Cross Market dataset~\cite{Ren2019dataset} consists of network traffic from 229 apps randomly selected from the most popular Android apps in three countries (China, India, and USA) in 2017~\footnote{At the time this research was conducted, the Cross Market Dataset was publicly available. However, as of the publication date, the dataset is no longer accessible to the public.}. The MAppGraph dataset~\cite{Pham2021mappgraph} which was collected in 2021 and available upon request. The original dataset has traces from 101 popular Android apps in Vietnam; however, the version shared with researchers includes 81 apps. Lastly, the dataset by Mankowski et al.~\cite{Mankowski2023dataset} from 2023 comprises traces for 90 Android apps from the German market. The traces from the Cross Market and Mankowski datasets include one traces per app with an average duration of 5 minutes. In contrast the MAppGraph dataset has 330 minutes of traffic per app on average, totaling nearly 500 GB of data. 

Therefore, we selected the MAppGraph dataset as our source for mobile app traffic traces to generate 1,000 synthetic user traces. This dataset was chosen because it provides data for 81 mobile apps, offering a broad range of app traffic. Each app includes an average of 330 minutes of traffic, which is more extensive compared to other available datasets. Furthermore, its collection in 2021 ensures that the traffic is both recent and relevant. These factors make the MAppGraph dataset ideal for creating diverse and reliable synthetic user traces.

\subsection{MAppGraph Dataset Description and preprocessing}  

The MAppGraph dataset comprises encrypted traffic captures generated by Android mobile apps. Data we collected at Tan Tao University in Vietnam during multiple sessions. In each session, volunteer students used smartphones provided by the research team to access apps from a predefined list. The primary goal was to record traffic from individual app executions by human users rather than complete user profiles. Despite human users generated the traffic, the dataset contains only records of app executions, not continuous user-specific traces.

\begin{figure*}[!ht]
  \centering
  \includegraphics[width=\textwidth]{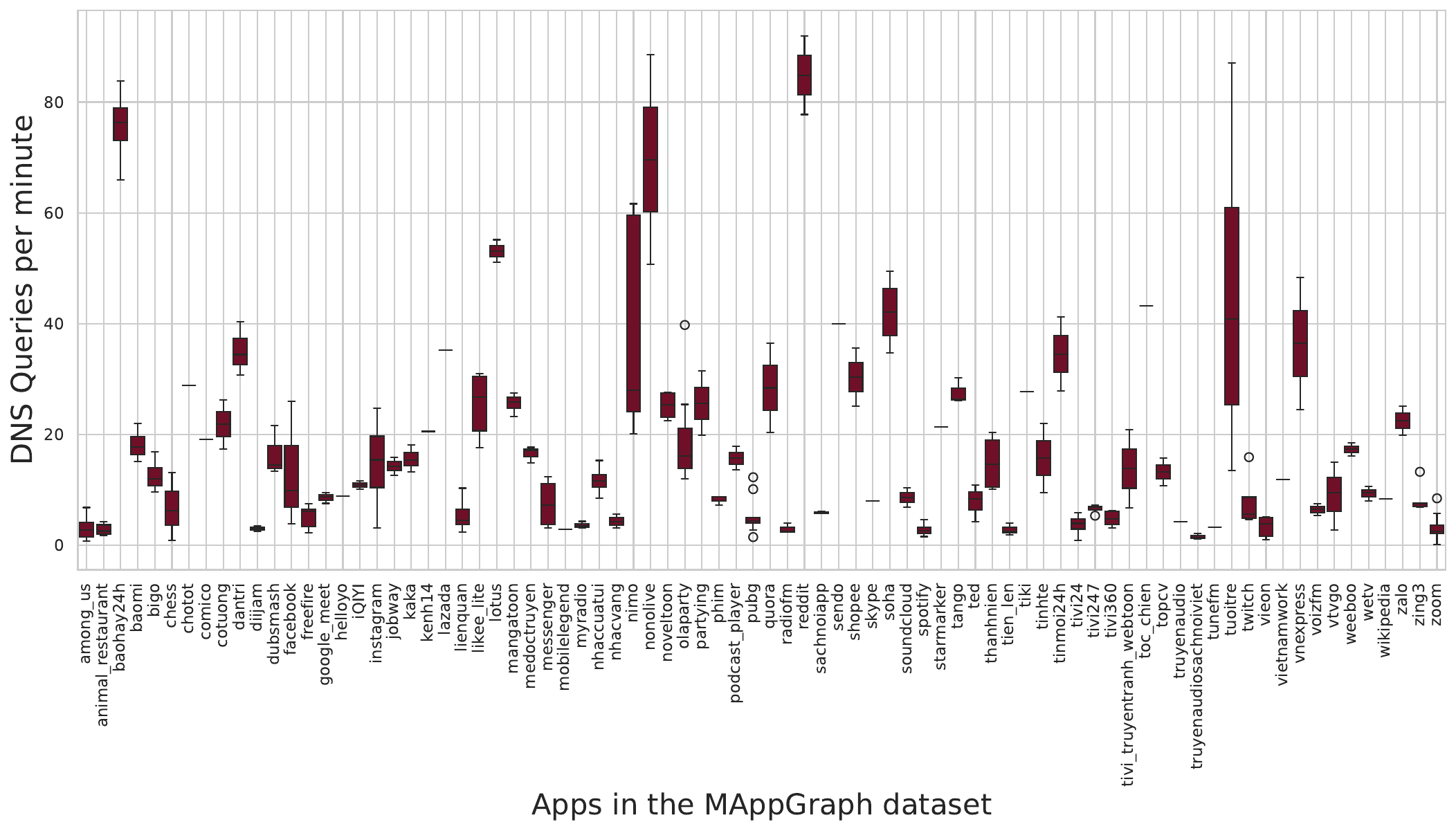} 
  \caption{DNS queries per minute in the MAppGraph dataset.}
  \label{fig:QueriesPerMinDATASSET}
\end{figure*}

Although the MAppGraph dataset primarily contains encrypted traffic, it also includes unencrypted DNS (Do53) traffic. In this study we use the DNS traffic to obtain a user profiling, as cleartext DNS queries can be used to infer the app generating the traffic~\cite{Campo2024dns}. The preprocessing step involved extracting the timestamp and domain name from the DNS requests of each PCAP file. This extracted parameters are stored in a CSV file per trace; in what follows, we will refer to this CSV files as the traces of the apps or users. In Fig.~\ref{fig:QueriesPerMinDATASSET} illustrates the range of DNS queries per minute for each app in the MAppGraph dataset. We can see that the density of DNS queries is different per app. 

\subsection{Generation of synthetic user traces}

Our approach for generating synthetic user traces involves a twofold process: 
first, we analyse the mobile app usage behavior in order to obtain the distribution of the installation percentage of the categories and apps. Then, we create the synthetic user traces and we assign traffic to them. The code to create 1000 synthetic user traces is available in GitHub\footnote{https://github.com/AndreaJimBerenguel/user\_profiling}. 

\subsubsection{Mobile App Usage Behaviour}

Previous studies have shown that mobile app usage follows a power-law distribution: users tend to rely on one main app, and the likelihood of using additional apps drops off according to a power law~\cite{Siebers2023powerlaw}. 

To capture this behavior, we analyze the number of installations for each app present exclusively in our dataset. Detailed installation numbers for each app, which can be found in~\cite{Pham2021mappgraph}, support our analysis and demonstrate that the installation numbers follow a power-law trend. Then, we grouped the apps into 10 categories based on the app-category of Google play. In Table~\ref{tab:apps_categories} are presented the categories and the number of apps per category from the MAppGraph dataset. The numbers of apps sum 80 because we omitted one app due to insufficient traffic trace data. 

\begin{table}[ht]
    \centering
    \begin{tabular}{@{}l c@{}}
        \toprule
        \textbf{Category} & \textbf{Number of apps} \\
        \midrule
        BOOKS AND REFERENCE  & 9  \\
        BUSINESS             & 4  \\
        COMMUNICATION        & 4  \\
        EDUCATION            & 2  \\
        ENTERTAINMENT        & 15 \\
        GAMES                & 10 \\
        MUSIC AND AUDIO      & 10 \\
        NEWS AND MAGAZINES   & 11 \\
        SHOPPING             & 5  \\
        SOCIAL               & 10 \\
        \bottomrule
    \end{tabular}
    \caption{Categories and number of apps per category.}
    \label{tab:apps_categories}
\end{table}

\begin{figure}[ht]
    \centering
    \subfigure[Installation probabilities of the apps aggregated by categories (probabilities inter-categories)]{
        \includegraphics[width=\columnwidth]{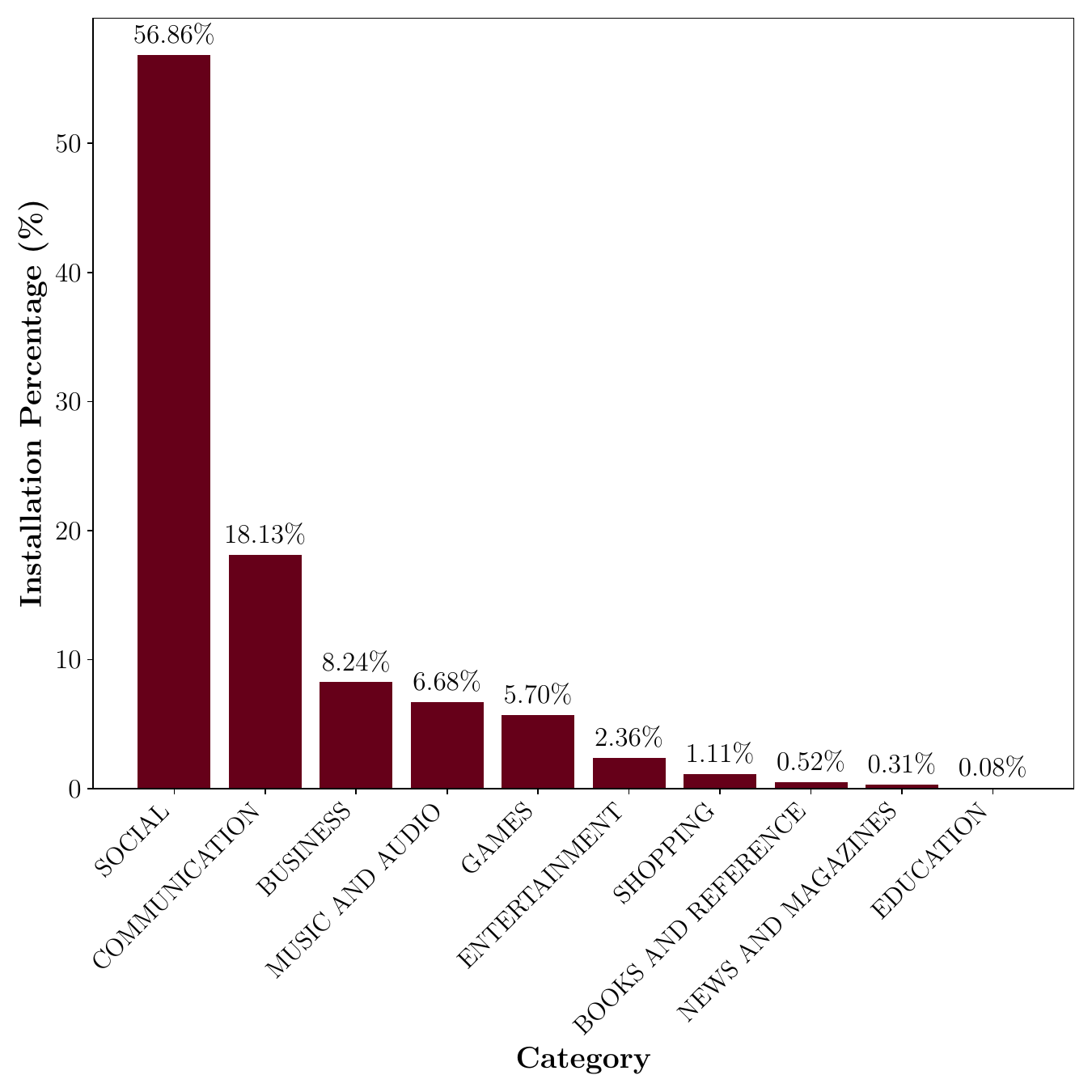}         \label{fig:histogram_prob_installation_per_categories_new_categories}
    }
    \hfill
    \subfigure[Installation percentage of the apps from \textit{games} (probabilities intra-categories)]{
        \includegraphics[width=\columnwidth]{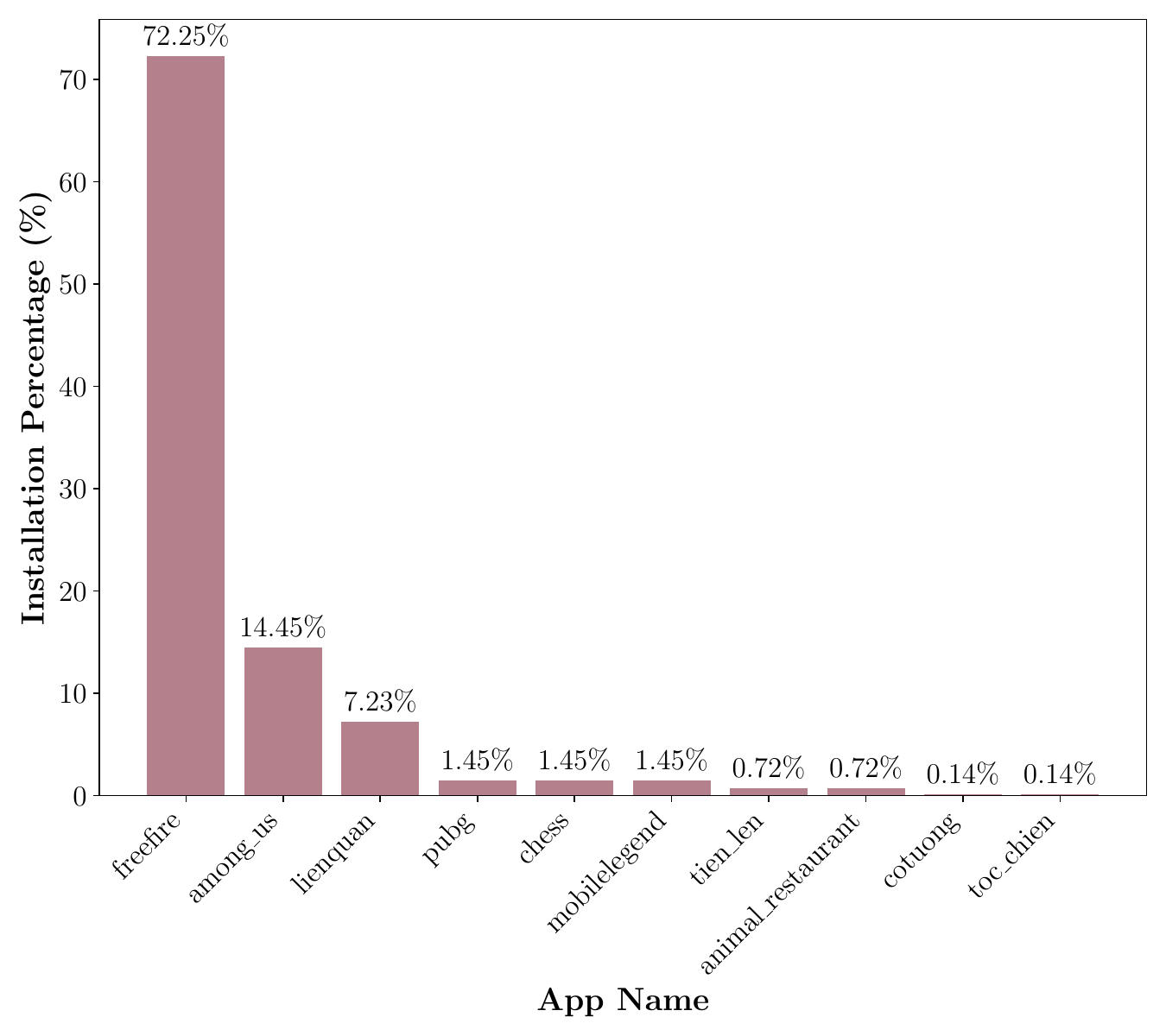}         \label{fig:histogram_prob_installation_games_3}
    }
    \caption{Histograms illustrating the installation probabilities inter and intra-category which follow a power-law distribution.}
    \label{fig:histogram_prob_installation}
\end{figure}

The Fig.~\ref{fig:histogram_prob_installation} are represented the probabilities of installation: the histogram shown in Fig.~\ref{fig:histogram_prob_installation_per_categories_new_categories} 
illustrates that the installation probabilities aggregated by categories (inter-category) follow a power-law distribution. Moreover, in Fig.~\ref{fig:histogram_prob_installation_games_3} are represented the individual installation probabilities of the apps from the category~\textit{games} as an example. We can see that when we analyze the apps within the same category (intra-category), the installation probabilities of individual apps within a category also exhibit a similar power-law trend.

Based on the power-law distribution of the inter-category and intra-category app usage we obtain a set of apps for each user. In our methodology, we assume that the probability of a user selecting a particular app is proportional to its installation probability. 
\paragraph{Category and weight assignation} First, we assign to each user the categories and its weight. In this case, the weights are equivalent to usage probabilities. 
We obtain samples from the power-law distribution of the categories. Each user gets a sample of this distribution. Because of the nature of the power-law distribution sampling, each user ends up with a unique weights vector, some categories may receive a zero weight while others receive a non-zero value. For example, user A obtains this categories and weights [`social: 68'; `communications: 11'; `business: 6'; `music and audio': 7; `games: 4';`entertainment: 4'; `shopping: 0'; `books and references: 0'; `news and magazines: 0'; `education: 0'].

\paragraph{App per category} Next, for each category that has a non-zero probability, we randomly select one app from that category. This selection is not uniform but weighted according to the power-law distribution observed among the apps within that category. The end result is a vector named \textit{user app usage} for each user that maps each selected category and app to its assigned percentage of usage. For example, we assign to user A one app per category with a non-zero value. In Table~\ref{tab:apps_userA} we present the assignation of apps per category to user A. 


\begin{table}[ht]
    \centering
    \begin{tabular}{@{}l l c@{}}
        \toprule
        \textbf{Category} & \textbf{App} & \textbf{Weight} \\
        \midrule
        Social             & Facebook    & 68 \\
        Communications     & Messenger   & 11 \\
        Business           & Jobway      & 6  \\
        Music and Audio    & SoundCloud  & 7  \\
        Games              & Freefire    & 4  \\
        Entertainment      & Nimo        & 4  \\
        Shopping           & --         & 0  \\
        Books and References & --       & 0  \\
        News and Magazines & --         & 0  \\
        Education          & --         & 0  \\
        \bottomrule
    \end{tabular}
    \caption{Apps and Weights per Category for User A}
    \label{tab:apps_userA}
\end{table}

\subsubsection{Synthetic User Trace Generation and Traffic Assignment}

\begin{figure*}[ht]
    \centering
    \subfigure[Representation of the \textit{user app usage} vector of User A and User B.]{
        \includegraphics[width=0.48\textwidth]{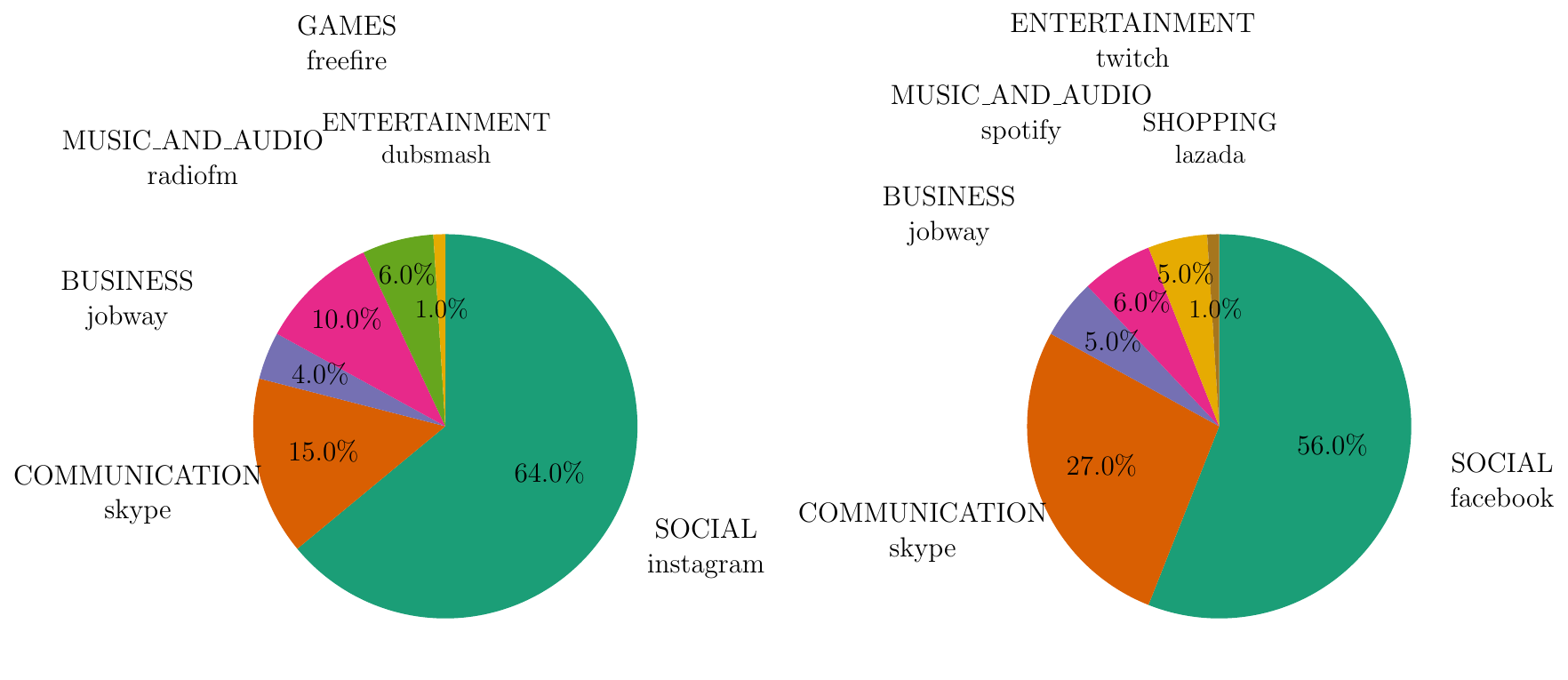}         \label{fig:latex_piechart_user_profiles_new_categories_all_correct_2}
    }
    \hfill
    \subfigure[Representation of the \textit{user DNS profiling} vector of User A and User B.]{
        \includegraphics[width=0.48\textwidth]{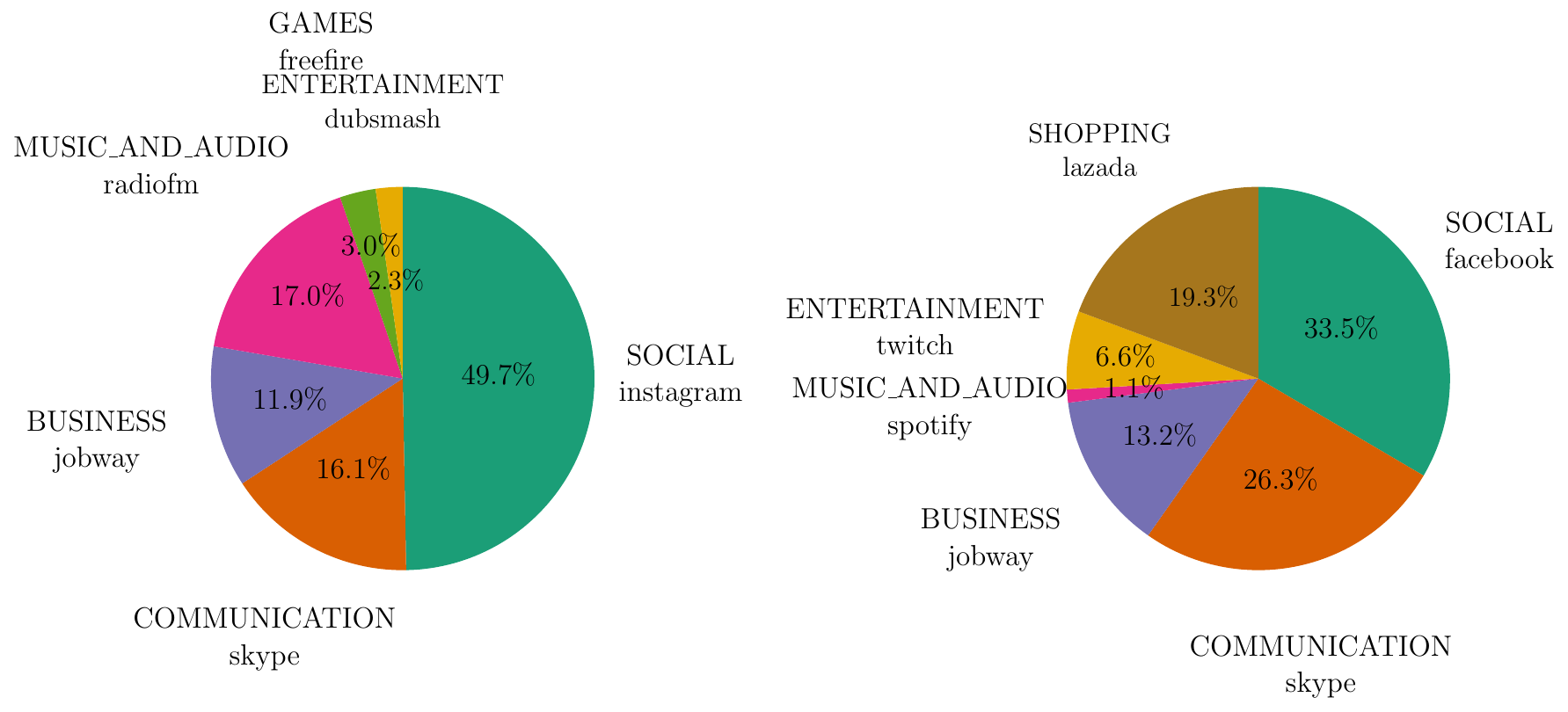}         \label{fig:latex_piechart_user_profiling_by_percentage_DNStraffic_new_categories_all_correct_2}
    }
    \caption{Comparison between \textit{user app usage} vector and \textit{user DNS profiling} vector of User A and User B.}
    \label{fig:piecharts}
\end{figure*}

We generated traces for 1,000 synthetic users. First, for each user we defined a \textit{user app usage} vector that includes the app categories, selected apps, and the corresponding percentage of usage for each app, as described in the previous subsection. Fig.~\ref{fig:latex_piechart_user_profiles_new_categories_all_correct_2} shows an example of the \textit{user app usage} vectors of User A and User B. The diagrams represent their respective categories, apps, and usage percentages derived from the power-law distribution. Using these \textit{user app usage} vectors, we then assigned traffic traces from the corresponding apps to each user, drawing from the traffic available in the MAppGraph dataset.

The synthetic trace for each user consists of 100 minutes of observed traffic. Within this fixed time window, we calculate the traffic interval for each app based on its usage percentage. We then extract the corresponding interval from the available traffic traces in the MAppGraph dataset. Although the power-law sampling naturally produces varied segments, we further enhance variability by selecting different starting points for each extraction. This prevents identical traffic patterns across users. The resulting traffic trace for each user is labeled by app and stored in a CSV file. For example, as shown in Fig.~\ref{fig:latex_piechart_user_profiles_new_categories_all_correct_2}, user A has a 100-minute trace comprising 68 minutes of Facebook traffic, 11 minutes of Messenger traffic, 6 minutes of Jobway traffic, 7 minutes of SoundCloud traffic, 4 minutes of Freefire traffic, and 4 minute of Nimo traffic.

\subsection{User Profiling}


After generating synthetic user traces, we apply the user profiling described in Section~\ref{sec:Formal Problem Statement}. We profile the users based on the percentage of DNS traffic generated by each app within its respective time interval. This method builds a user profile by focusing on the specific DNS traffic produced by each app. Unlike previous approaches as~\cite{Shaman2019userprofiling} that derive profiles from generic network traffic, our method leverages traffic traces that are directly labeled with the app identity.

Each user trace is stored in a CSV file, we stored each timestamp and domain name of the DNS requests labeled per app. For each user, we calculate the percentage of DNS traffic contributed by each app during the observation period. These percentages are then mapped to their corresponding app categories. The final output for each user is a vector named \textit{user DNS profiling}, equivalent to the vector \textit{user app usage}, where each element represents the percentage of DNS traffic for a specific category. 

In Fig.~\ref{fig:latex_piechart_user_profiling_by_percentage_DNStraffic_new_categories_all_correct_2} shows the \textit{user DNS profiling} for Users A and B, as illustrated in the previous example. Our approach effectively captures the user’s behavior based on the DNS traffic generated by the apps. It is important to note that the observed DNS query percentage does not directly reflect actual app usage time; different apps generate DNS queries at different rates. For example, we can appreciate in Fig.~\ref{fig:piecharts} between both pie charts, one app might generate many queries in a short period as seen with User A in the category \textit{entertainment}, while another app produces fewer queries over a longer time as it happens in the category \textit{music and audio} in User B.

\begin{figure}[ht]
  \centering
  \includegraphics[width=\columnwidth]{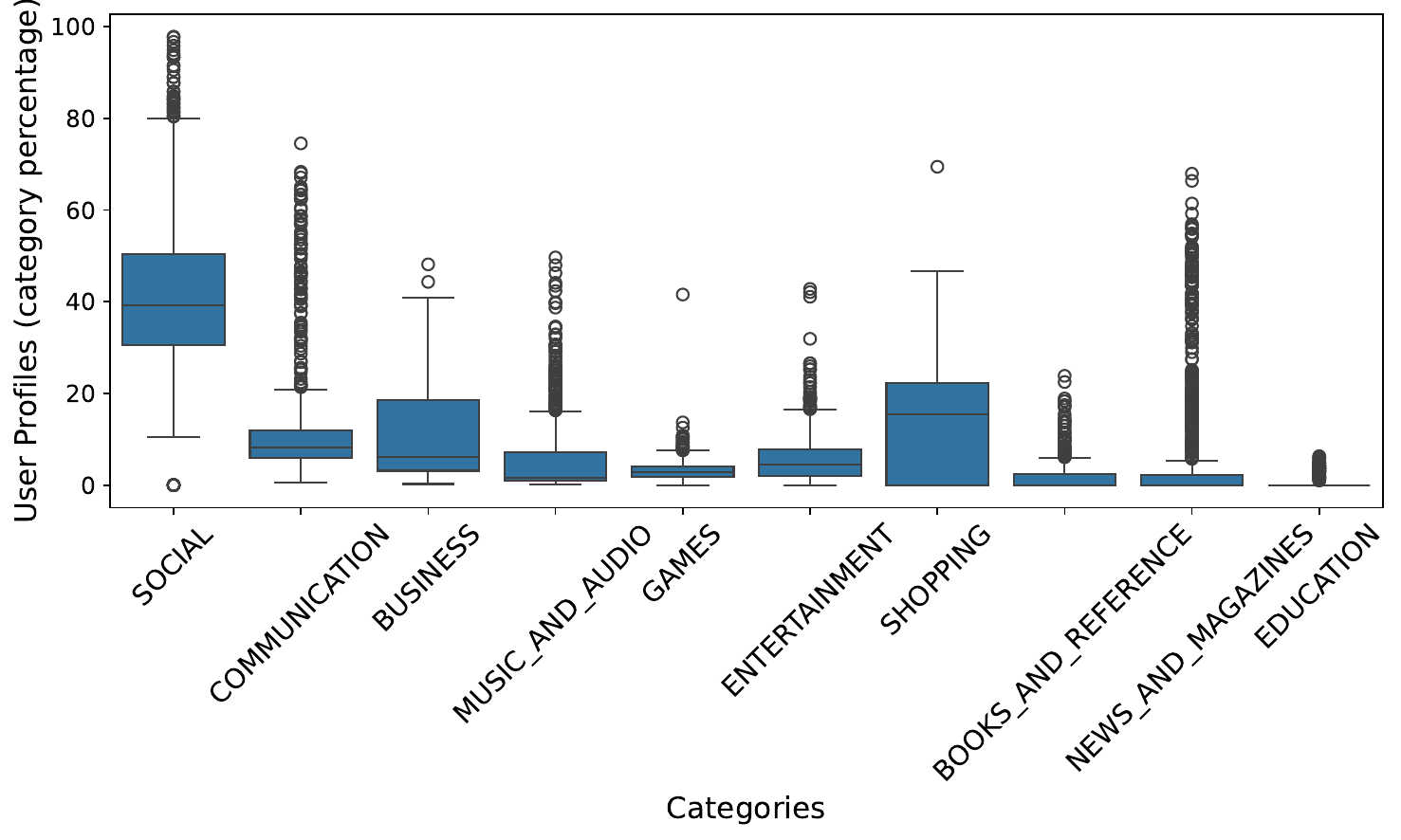} 
  \caption{Representation of the average percentage of DNS traffic per app for each user}
  \label{fig:boxp_user_profiling_by_percentage_DNStraffic_new_categories}
\end{figure}

In addition, we analyze the variability of the user profiles obtained by our method. Fig.\ref{fig:boxp_user_profiling_by_percentage_DNStraffic_new_categories}
displays the average percentage of DNS traffic per app for each user. While we anticipated a high percentage of DNS traffic in the \textit{social} category, our results also reveal that the \textit{business}, \textit{entertainment}, and \textit{shopping} categories exhibit high percentages, even though the corresponding time intervals are much shorter. This observation suggests that apps in these categories have a higher rate of DNS queries compared to other categories.

\section{Evaluation}
\label{sec:Evaluation}
\noindent

In this section, we analyze the extent to which our proposal can help users protect their privacy when using mobile apps in a real-world scenario. At the same time, we assess the impact of generating fake DNS queries on the quality of the DNS resolution service using the utility metric we define as the forgery rate $\rho$.

\subsection{Results}
\label{sec:Evaluation:Results}
\noindent

To evaluate how effectively our proposal enhances user privacy protection, we designed an experiment that measures the individual privacy improvement for each synthetic user in the dataset described in Section~\ref{sec:Users_Dataset}.
We compute the privacy risk function for each user based on the perturbation rate $\rho$ for the three DNS query forgery strategies defined in Section~\ref{sec:DNS Query Forgery Against Profiling:Metrics}, namely, Uniform, TrackMeNot-based, and Optimized. Ultimately, we compile all collected data and represent the privacy gain as a function of  $\rho$ musing the 10\%, 50\%, and 90\% percentiles of the two best-performing strategies.

\begin{figure*}[tb]
	\centering	
		\subfigure[10th percentile - Entropy-based]
		{\includegraphics[width=0.30\textwidth]{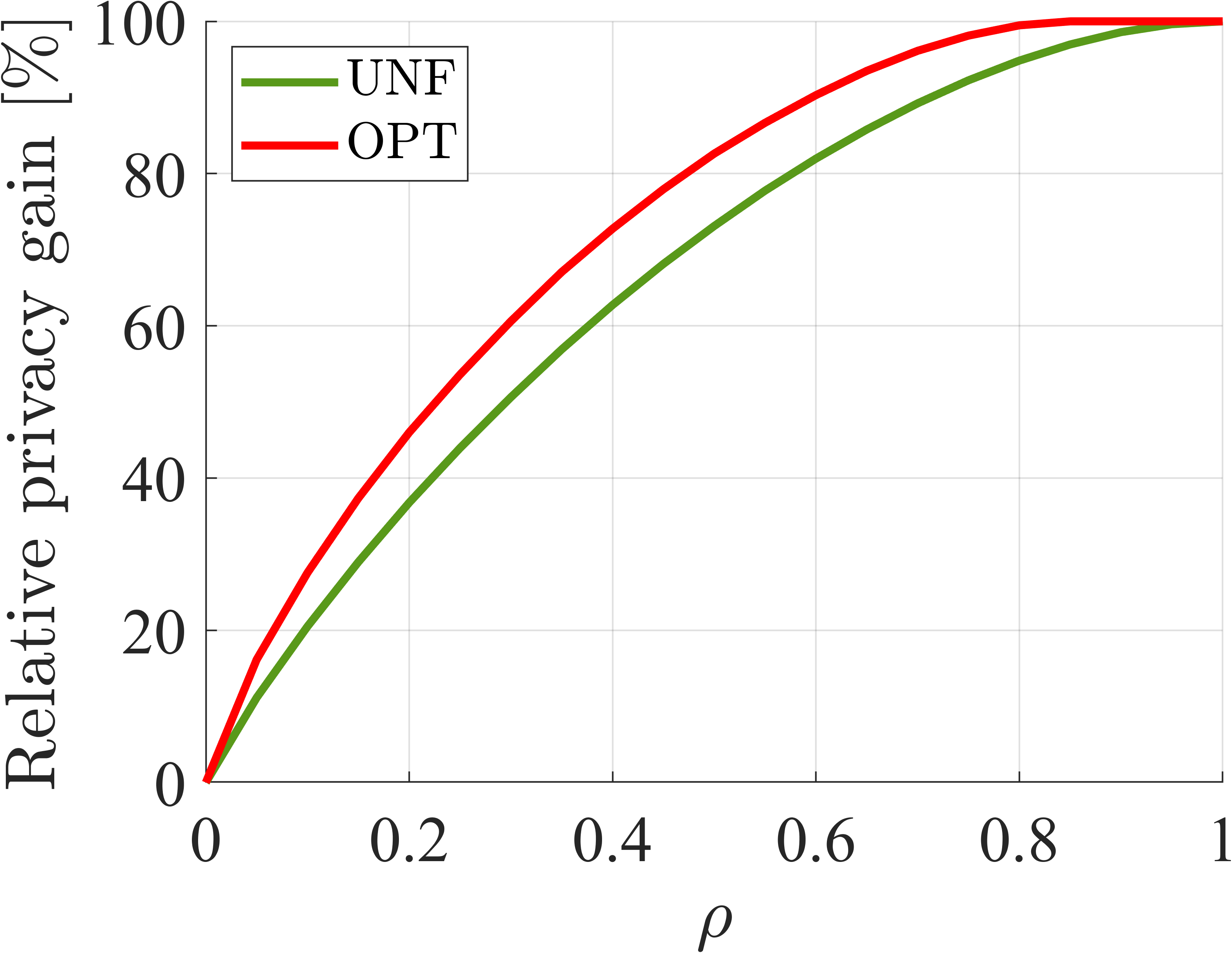}} 
		\subfigure[50th percentile - Entropy-based]
		{\includegraphics[width=0.30\textwidth]{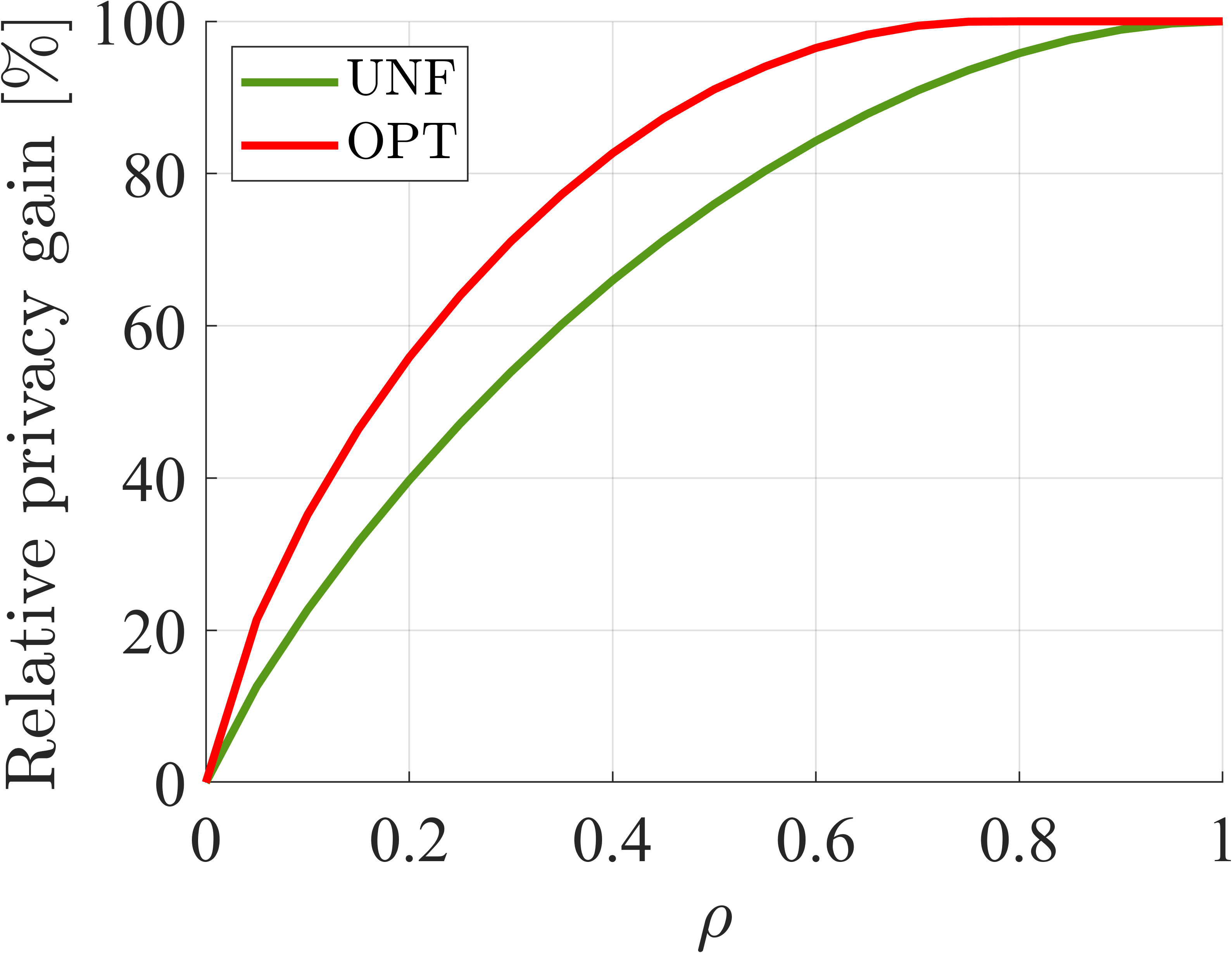}}
		\subfigure[90th percentile - Entropy-based]	
		{\includegraphics[width=0.30\textwidth]{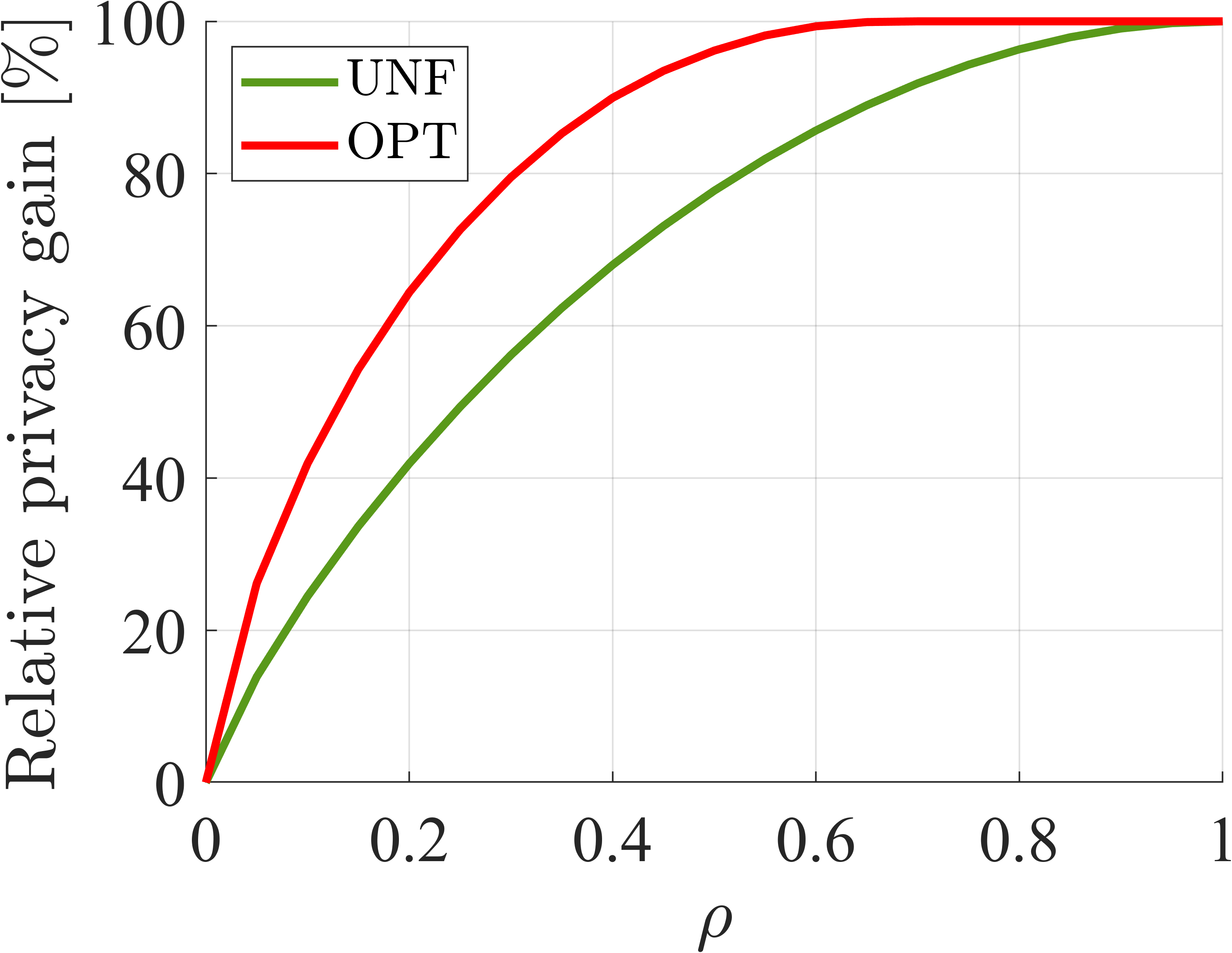}}

		\subfigure[10th percentile - Divergence-based]
		{\includegraphics[width=0.30\textwidth]{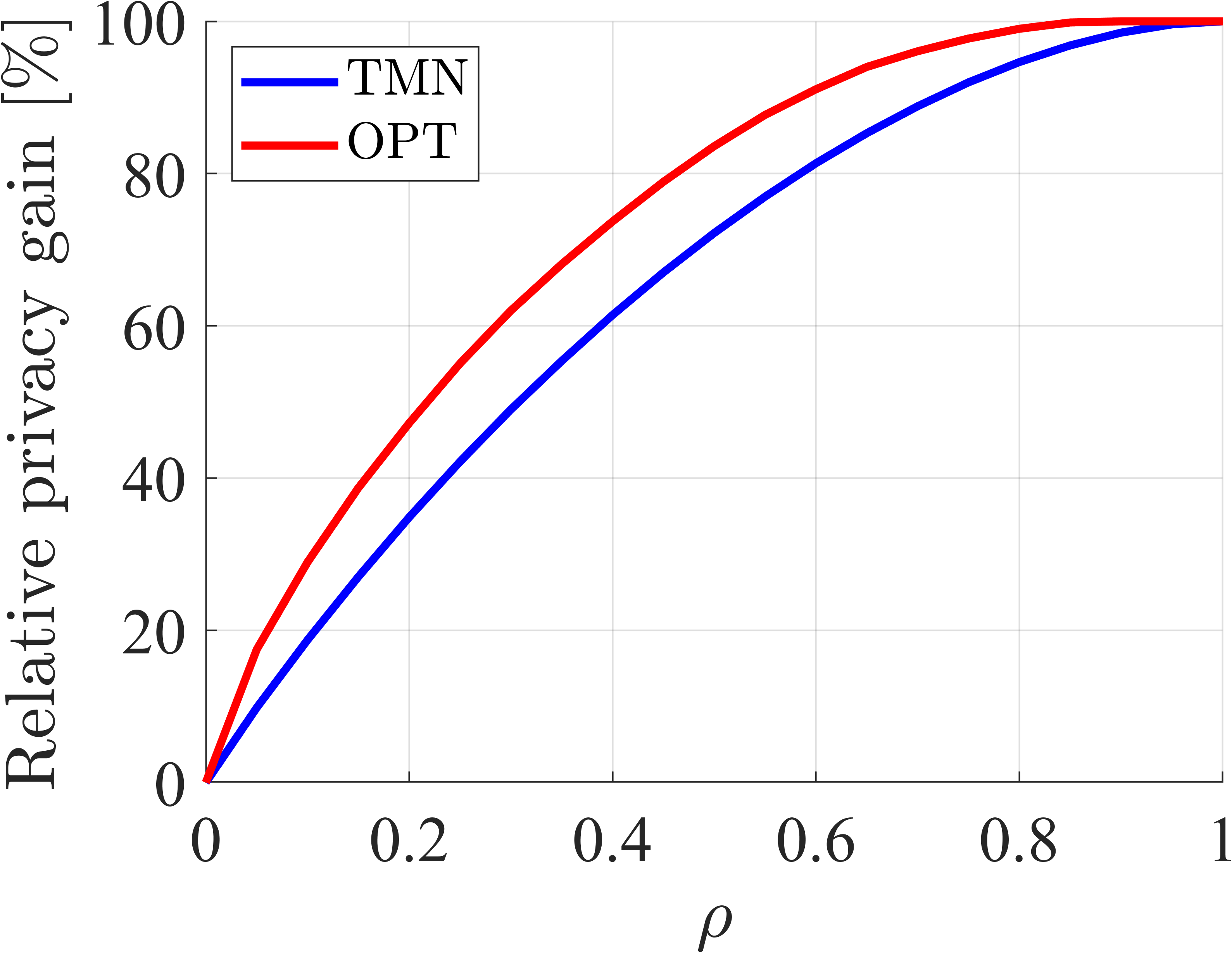}} 
		\subfigure[50th percentile - Divergence-based]
		{\includegraphics[width=0.30\textwidth]{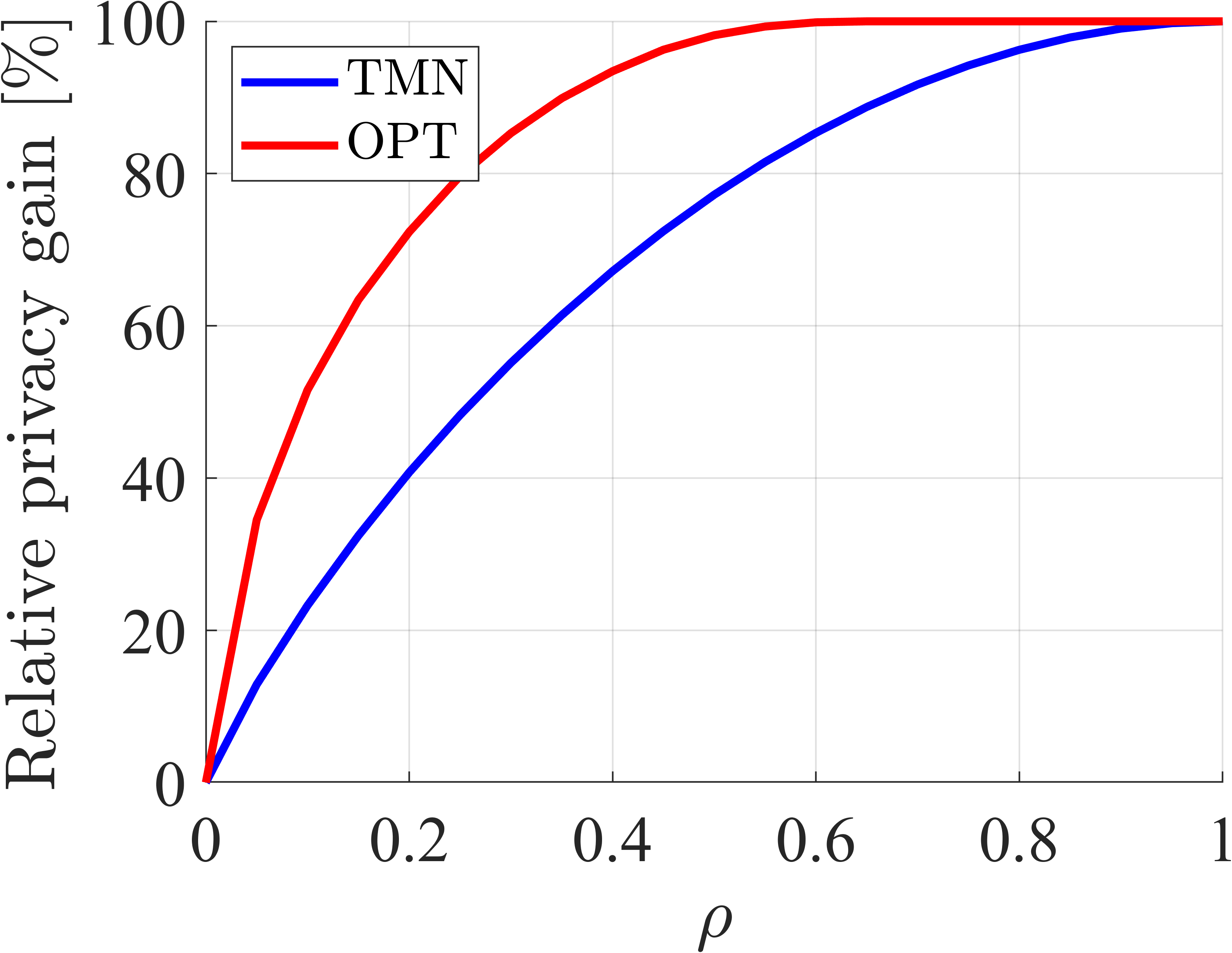}}
		\subfigure[90th percentile - Divergence-based]	
		{\includegraphics[width=0.30\textwidth]{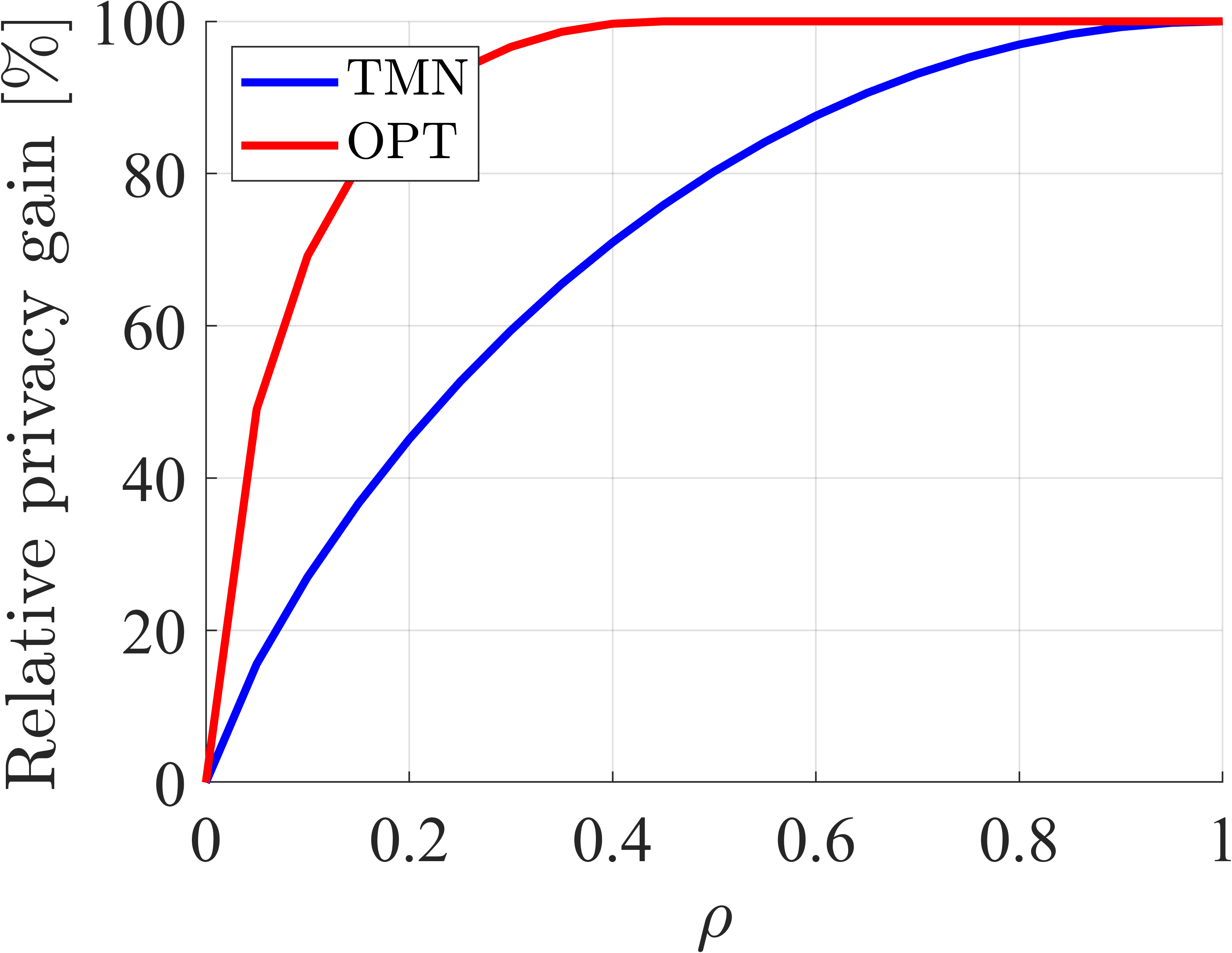}}

	\caption{The representation of the relative privacy gain for the two best forgery strategies as a function of the perturbation rate $\rho$, across the entropy and divergence and the 10\%, 50\%, and 90\% percentiles.}
	\label{fig:mappgraph_dataset_percentiles}
\end{figure*}

In Fig.~\ref{fig:mappgraph_dataset_percentiles}, we present the relative privacy gain for the two best forgery strategies as a function of the perturbation rate $\rho$, conveniently partitioned into 21 values, across the two privacy metrics and three percentiles. Undoubtedly, in all cases, the optimized strategy outperforms the suboptimal ones. With the optimized mechanism, we achieve a 100\% privacy improvement for 90\% of users with false query rates above 60\% when the risk metric is based on entropy (i.e., when no reference profile is available, and the uniform distribution is used). When the risk is based on \ac{KL} divergence, the same improvement is obtained at perturbation rates above 40\%. However, a 50\% improvement for most users is only achievable with perturbation rates below 20\%. These parameters guide us in understanding the traffic overhead users must assume to achieve reasonable privacy gains with an optimized mechanism. In contrast, suboptimal mechanisms require a full perturbation rate to attain any 100\% improvement.

\begin{figure}[h]
	\centering
	\includegraphics[scale=0.55]{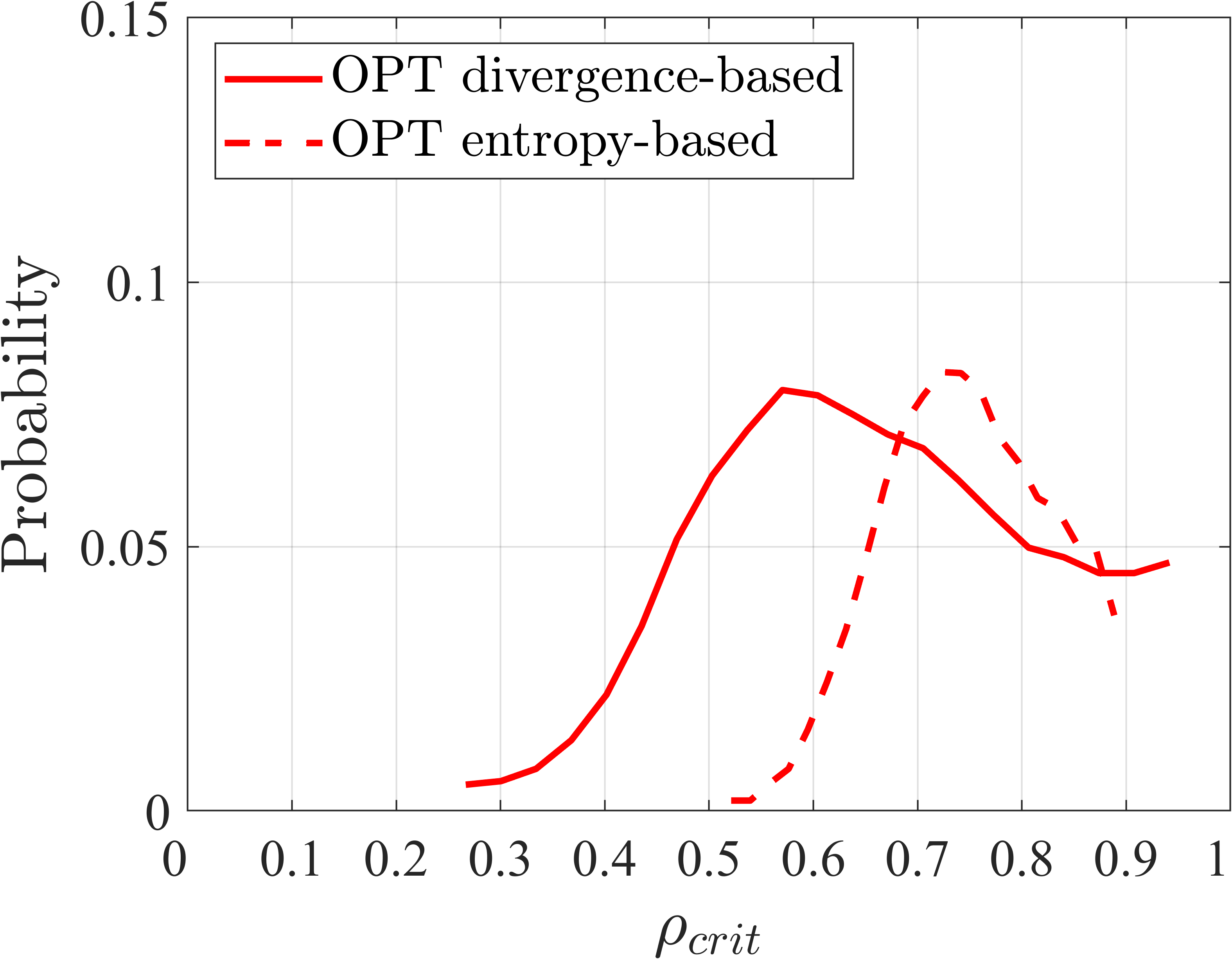}
	\caption{The distribution of critical perturbation rate values for the 1,000 users in our dataset.}
    \label{fig:mappgraph_dataset_critical_rate_opt_stretegy}
\end{figure}

Furthermore, for the optimized DNS query forgery strategy, Fig.~\ref{fig:mappgraph_dataset_critical_rate_opt_stretegy} displays the distribution of critical perturbation rate values for the 1,000 users in our dataset. The conclusions drawn from these results further reinforce those obtained from the relative privacy gain analysis.

\subsection{discussion} 
\label{sec:Evaluation:Discussion}
\noindent

In this section, we discuss the results obtained from our DNS privacy model and compare our approach with state-of-the-art DNS privacy models mentioned in Section~\ref{sec:State of the Art:Applications of PETs to DNS}.

Our DNS query forgery approach performs best with the Optimized strategy, achieving 100\% privacy improvement for 90\% of users at false query rates above 60\% with entropy-based metrics and above 40\% with \ac{KL} divergence metrics. A 50\% privacy improvement requires perturbation rates below 20\%. These findings clarify the traffic overhead needed for effective privacy protection, while suboptimal mechanisms require 100\% perturbation for similar results.

Despite the overhead our model introduces, it offers significant advantages over other DNS privacy approaches. Unlike PIR-based DNS privacy proposals like those presented in~\cite{Bhat2019pir,Zhou2024pir}, our model can adapt to a dynamically changing DNS environment. PIR-based approaches require static tables and server-side implementation, meaning both client and server must implement the privacy model for it to function properly.

When compared to collaborative DNS privacy models like \ac{NQA}~\cite{Arana2021nqa}, our approach offers distinct advantages. While \ac{NQA} can also adapt to DNS dynamism, it depends on creating a trust ecosystem with other users. In contrast, our model follows a zero-trust approach where any third party is considered a potential privacy threat. This provides user-side privacy without requiring trust in external entities.

More recent developments like \ac{ODNS}~\cite{Schmitt2019odns} or \ac{ODoH}~\cite{rfc9230} require implementation of the privacy model both at the user side (stub resolver) and at the DNS server side (dedicated \ac{ODNS} resolver). They also introduce computational overhead for encrypting each query and forwarding it to a dedicated \ac{ODNS} server for decryption before the DNS server can respond to the request.

Our model, while adding false queries to the traffic, offers several key benefits: it does not interfere with app functionality, does not require third parties to guarantee user security (following a zero-trust model), and provides user-side privacy. The trade-off is a controlled increase in network traffic, which our results show can be optimized to balance privacy gains and performance impact.





\section{Practical Adaptation of Query Forgery for Mobile Apps}
\label{sec:Practical Adaptation of Query Forgery for Mobile Apps}
\noindent

We consider our proposal to be feasible for real-world implementation. With this premise in mind, we dedicate this section to establishing the foundations for the eventual deployment of a system that enables mobile app users to protect their privacy by perturbing the DNS traffic generated during their online activity. To achieve this, we adopt a high-level modular scheme, in which different modules perform specific functions within the system and interact with each other, as well as externally, to achieve the defined objective.

In practical terms, we envision a mobile app or a similar tool installed on the user's device, functioning as a decision-support system. That is, the app generally operates in the background and, upon detecting a privacy threat or compromise, alerts the user and presents possible countermeasures, allowing them to choose the rate of false DNS queries or the perturbation strategy itself. It is important to recall the principle underlying \ac{DPT}—\textit{hard privacy}—where the user is responsible for their own privacy, without relying on potentially untrustworthy third parties.

Before detailing the main functional components of our design, we must specify how a user's profile could be obtained locally in an app implementing our technique. To this end, we base our approach on three assumptions regarding the user profile.

First, as a common knowledge hypothesis, we assume that both entities—the mobile app and potential privacy attackers—operate over an identical set of interest categories. Consequently, based on their respective categorization algorithms, they derive the same user profile. This assumption holds as long as these categories belong to standardized sets available to both parties.

Second, in terms of profile initialization, we assume that in order to determine whether to add traffic to a specific category, our approach requires an initial user profile. One possible way to address this is by establishing a training phase prior to deployment.

Additionally, we assume the concept of a long-term profile, meaning that the user's profile does not change frequently, in line with~\cite{gauch2007user}. The profile stabilizes after the initialization phase, once the user has shared a significant number of elements. However, we acknowledge that, in practice, user interests may vary significantly over time. Therefore, our implementation should account for this dynamic aspect.
 
\begin{figure}
	\centering
	\includegraphics[scale=0.35]{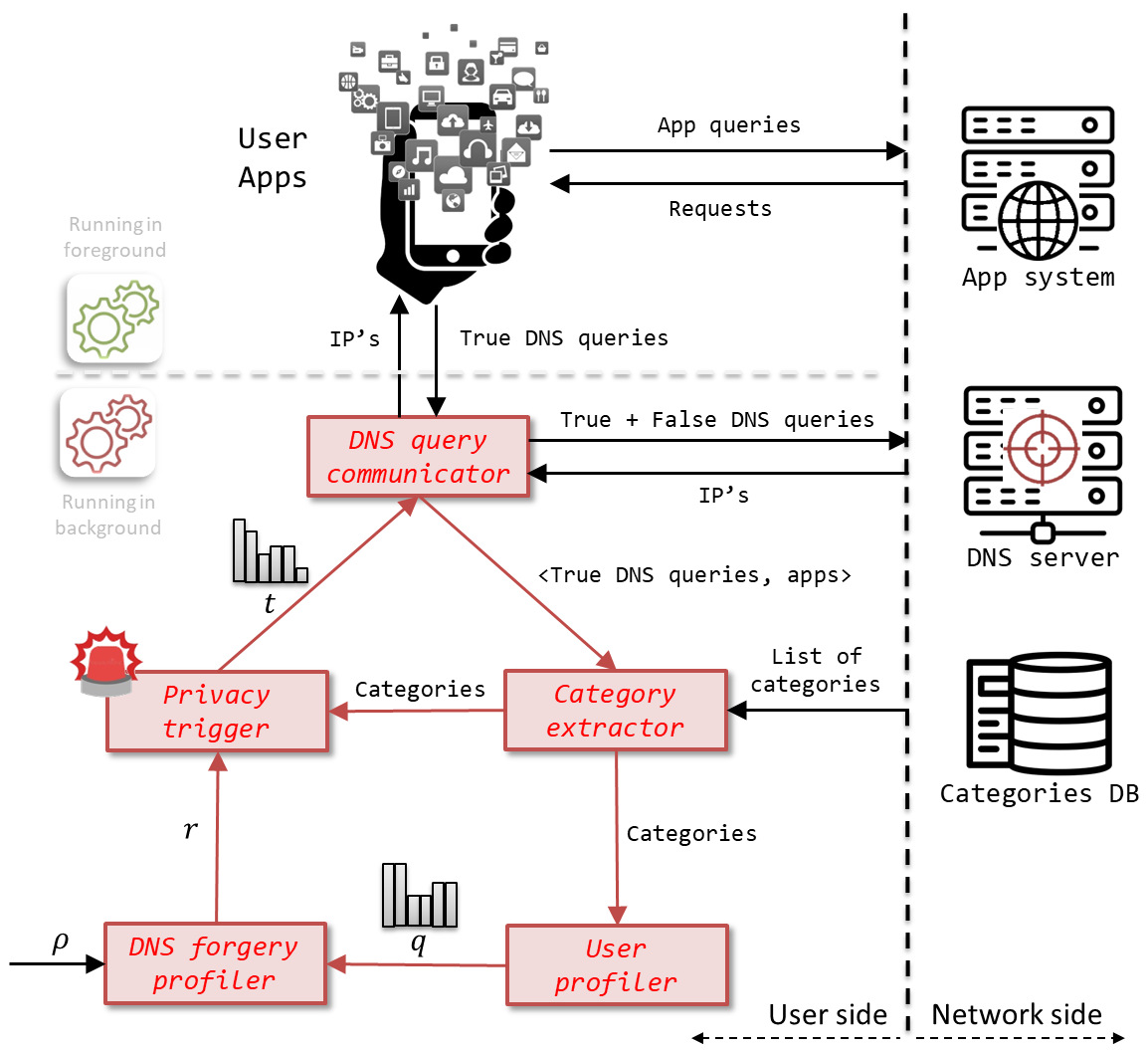}
	\caption{Software architecture for an implementation of our privacy proposal in the mobile app scenario.}
	\label{fig:architecture}
\end{figure}

Fig.~\ref{fig:architecture} illustrates a modular architecture for a hypothetical implementation of our methodology as a DNS‐query forger. It consists of a series of modules that interact locally and/or with the system, each performing a specific function based on the parameters it receives. From a general perspective, the figure depicts a user interacting with a single, straightforward DNS server. This server provides the user with relevant information, specifically IP addresses for resolving queried domains, generating a communication flow composed of DNS traces that third parties could exploit in undesirable ways, compromising user privacy. The following section provides a functional description of the five modules that comprise the architecture.

The first module, and the only one interacting externally in our false DNS query app, is the \texttt{DNS Query Communicator}. This module is responsible for sending all DNS queries it receives, whether genuine—originating from the user's apps—or false, as dictated by our app’s results. Similarly, it receives the IP addresses resolved by the DNS server and returns them to the corresponding apps, provided they originate from a genuine query. Internally, the communicator forwards genuine DNS queries, along with the originating app, to the category extraction module for further processing. 

The \texttt{Categories Extractor} module plays a crucial role in user profiling. Based on DNS queries and their originating apps, this extractor processes the information primarily by mapping apps to their corresponding categories. By continuously updating its data, which can be modeled as a connection to an external database specializing in this type of information, this module determines the interest category associated with each DNS query—an essential step for constructing the user profile. The extracted category is then transmitted to both the user profiling module and the privacy trigger module.

The \texttt{User Profiler} module is responsible for generating and/or updating the user's real profile, denoted as $q$, based on the categories received from the category extractor module. We assume that the discrete histogram of relative frequencies constructed by this module stabilizes after a certain period. Consequently, profile initialization is a key aspect of our implementation, and we propose several initialization methods. For example, following~\cite{viejo2012using}, the profile can be initialized as $q = (0,\ldots,0), q \in \mathbb{R}^n$. Another alternative, based on the principle of maximum entropy, involves initializing the real profile as the uniform profile, $q = u$. A further option is to use a self-declared profile provided by the user, which, while not necessarily matching the profile inferred from their online activity, would eventually be replaced after the initial phase. Ultimately, the user's real profile $q$ is transferred to the DNS query forgery strategy module.

The core of our app is the \texttt{Strategy Generator} module, which ensures user privacy. This module implements various DNS query forgery strategies that we consider appropriate, such as those detailed in Section~\ref{sec:DNS Query Forgery Against Profiling:Metrics}. It generates corresponding distributions of false DNS queries, denoted as $r$, based on the perturbation percentage $\rho$ defined by the user. The output of this module is then passed to the privacy trigger module.

Finally, the \texttt{Privacy Trigger} module is responsible for alerting the user of potential privacy violations based on the distribution of false queries $r$. With probability $r_i$, an alert is issued regarding category $i$, allowing the user to decide whether to perturb their profile. The outcome of these actions is then transmitted to the communication module, which processes them accordingly, mixing false and genuine queries in the defined proportion to ultimately externalize the apparent profile $t$.  

\section{Conclusions and Future Work}
\label{sec:Conclusion}
\noindent

We have proposed DNS query forgery, a data perturbation strategy that minimizes personal information exposure in mobile app DNS traffic while following `hard privacy' principles. Our approach protects users from profiling by DNS resolvers without requiring trust in third parties.

Our evaluation demonstrates that query forgery effectively reduces profiling accuracy with minimal performance impact. We have quantified the trade-off between network overhead and privacy gains, showing that with optimal parameters, users can achieve significant privacy improvements at reasonable costs. The optimized query forgery strategy, in particular, delivers the best balance of privacy protection and efficiency.

We validated our DNS privacy model using a novel synthetic dataset of 1,000 users created from real mobile app traffic. This methodological innovation enables controlled experimentation on user profiling that would be difficult to achieve with real user data due to ethical and privacy constraints. Additionally, we proposed a modular software architecture that illustrates the feasibility of implementing our approach in real-world applications.


In future work, we consider exploring other user profiling techniques and delving deeper into the context of DNS traffic with targeted attacks on user privacy and new privacy-enhancing strategies.

\section*{Acknowledgment}
This work was supported by Grant COMPROMISE (PID2020-113795RB-C32 and PID2020-113795RB-C31) funded by MICIU/AEI/10.13039/501100011033, Grant QURSA (TED2021-130369B-C32) funded by MICIU/AEI/10.13039/501100011033 and European Union NextGenerationEU/PRTR, Grant MOBILYTICS (TED2021-129782B-I00) funded by MICIU/AEI/10.13039/501100011033 and European Union NextGenerationEU/PRTR, Grant SISCOM (2021 SGR 01413) funded by Generalitat de Catalunya through Agència de Gestió d'Ajuts Universitaris i de Recerca (AGAUR), Grant DISCOVERY (PID2023-148716OB-C33 and PID2023-148716OB-C32) funded by MICIU/AEI/10.13039/501100011033 and FEDER, UE, and from the I-Shaper Strategic Project (C114/23), due to the collaboration agreement signed between the Instituto Nacional de Ciberseguridad (INCIBE) and the Universidad Carlos III de Madrid, this initiative is being carried out within the framework of the Recovery, Transformation and Resilience Plan funds, funded by the European Union (Next Generation).

\bibliographystyle{unsrt}
\bibliography{references}

\phantomsection
\begin{IEEEbiography}[{\includegraphics[width=1in,height=1.25in,clip,keepaspectratio]{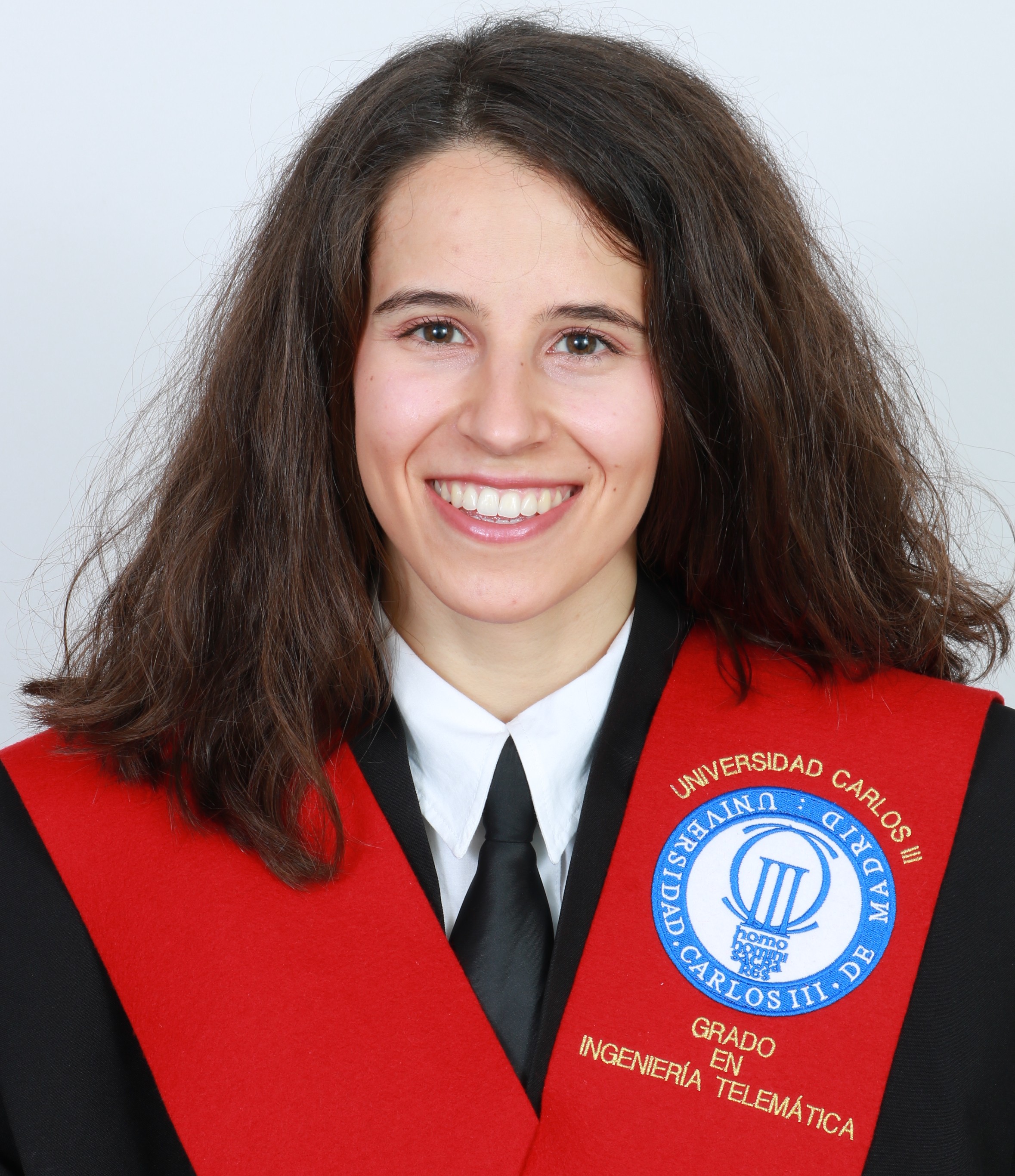}}]{Andrea Jimenez-Berenguel} is a PhD Student at the Department of Telematic Engineering of the Universidad Carlos III de Madrid. Her research journey is focused on Android traffic analysis and user privacy. She received her bachelor degree in Telematic Engineering in 2022 and her MS degree in Telecommunications Engineering in 2024, both from the Universidad Carlos III de Madrid.
\end{IEEEbiography}

\begin{IEEEbiographynophoto}{César Gil} received the bachelor's degree in statistics from the Universitat Politècnica de Catalunya (UPC), Barcelona, in 1997. He joined the UPC team in the SLOEGAT Project in 1998. Also received the MS degree in decision systems engineering from the Universidad Rey Juan Carlos (URJC) in 2012 and the MS degree in computational and mathematical engineering from the Universitat Oberta de Catalunya (UOC) and Universitat Rovira i Virgili (URV) in 2018. He is currently a Ph.D. candidate at UPC, where he investigates the optimal trade-off between privacy and data utility in personalized information systems.
\end{IEEEbiographynophoto}

\begin{IEEEbiography}[{\includegraphics[width=1in,height=1.25in,clip,keepaspectratio]{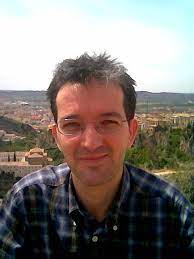}}]{Carlos Garcia-Rubio} received the Ph.D. degree from the Technical University of Madrid in 2000. He is an associate professor at the Department of Telematic Engineering of the University Carlos III of Madrid. His research focus is centered on mobile and wireless networked computing systems, and on the design and performance evaluation of communication protocols, mainly at the transport and application layers.
\end{IEEEbiography}

\begin{IEEEbiography}[{\includegraphics[width=1.3in,height=1.6in,clip,keepaspectratio, angle=-90]{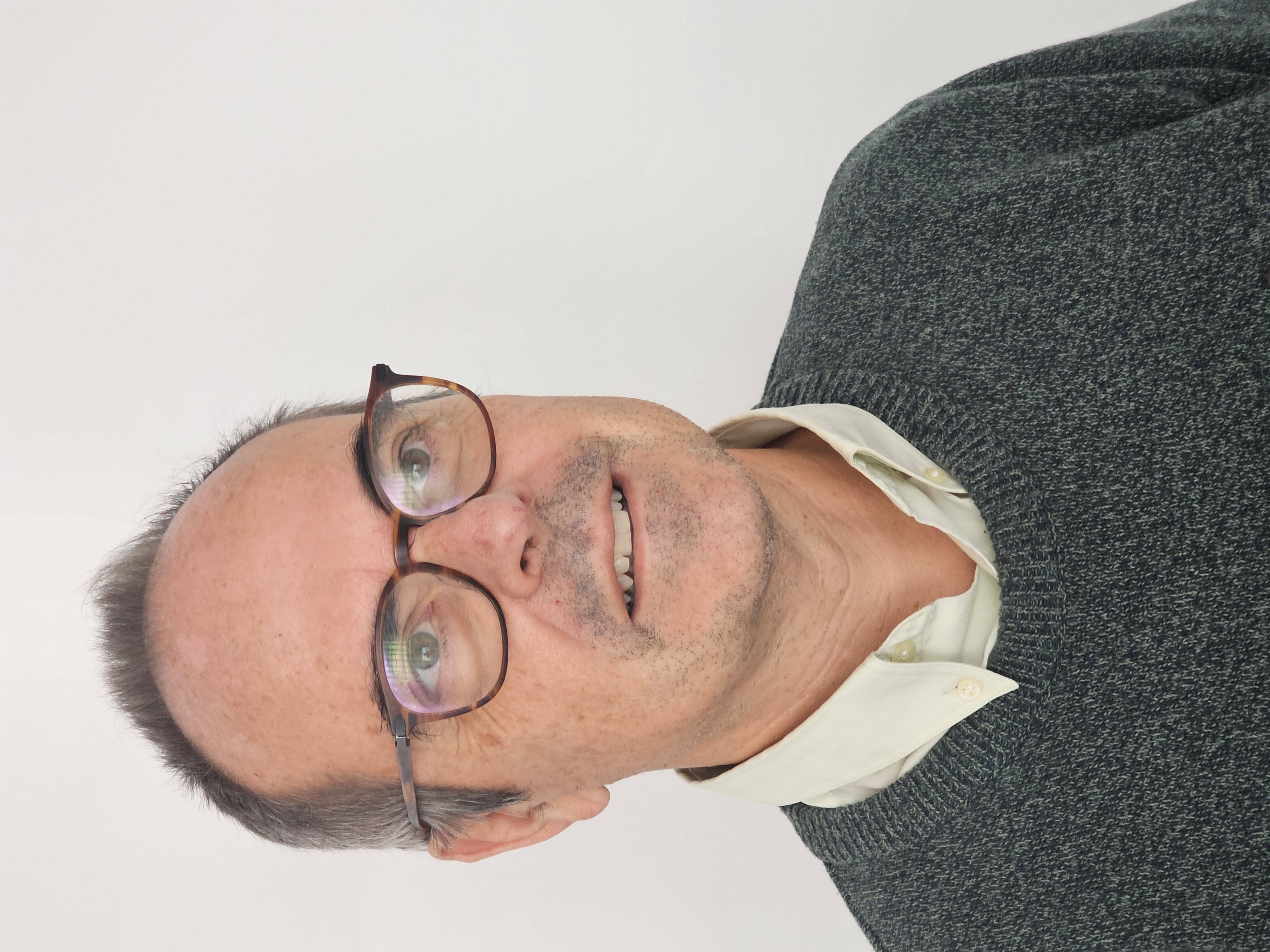}}]{Jordi Forné} received the M.S. and Ph.D. degrees in telecommunications engineering from Universitat Politècnica de Catalunya (UPC). Currently, he is a Full Professor with the Telecommunications Engineering School, Barcelona— ETSETB. He is also with the Smart Services for Information Systems and Communication Networks (SISCOM) Research Group, leading the research team on data privacy.
\end{IEEEbiography}

\begin{IEEEbiography}[{\includegraphics[width=1in,height=1.25in,clip,keepaspectratio]{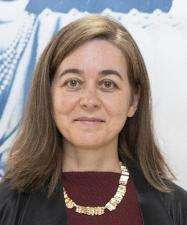}}]{Celeste Campo} received her Ph.D. degree from the University Carlos III of Madrid in 2004. She is an associate professor at the Department of Telematic Engineering of the University Carlos III of Madrid. Her research interests include design and performance evaluation of communication protocols for ad hoc networks, energy-aware communications, and middleware technologies for pervasive computing.  
\end{IEEEbiography}

\end{document}